%% file: EnergyAndSystemDependence.tex
\RequirePackage{lineno}
\documentclass[aps,prc,twocolumn,groupedaddress,showpacs,preprintnumbers,superscriptaddress,floatfix,amsmath,amssymb]{revtex4}
\usepackage [dvips]{graphics, color,epsfig}
\usepackage{xspace}
\usepackage{textcomp}
\usepackage{multirow}

\newcommand{\sNNtwohundred}{$\sqrt{s_{{NN}}}$ = 200 GeV\xspace}
\newcommand{\sNNsixtytwo}{$\sqrt{s_{{NN}}}$ =\xspace 62.4 GeV\xspace}
\newcommand{\sqrts}{$\sqrt{s}$\xspace}
\newcommand{\twohundred}{200 GeV\xspace}
\newcommand{\sixtytwo}{62 GeV\xspace}
\newcommand{\sNN}{$\sqrt{s_{NN}}$\xspace}
\newcommand{\stdassoc}{$1.5$~GeV/$c$ $<$ $p_T^{\mathrm{associated}}$ $<$ $p_T^{\mathrm{trigger}}$\xspace}
\newcommand{\stdtrig}{$3$~$<$ $p_T^{\mathrm{trigger}}$ $<$ $6$~GeV/$c$\xspace}
\newcommand{\trigrange}[2]{#1 $< p_T^{\mathrm{trigger}} <$ #2~GeV/$c$\xspace}

\newcommand{\assocrangevar}[1]{#1~GeV/$c$~$< p_T^{\mathrm{associated}} < p_T^{\mathrm{trigger}}$\xspace}
\newcommand{\pttrig}{$p_T^{\mathrm{trigger}}$\xspace}
\newcommand{\ptassoc}{$p_T^{\mathrm{associated}}$\xspace}
\newcommand{\npart}{$N_{\mathrm{part}}$\xspace}
\newcommand{\npartav}{$ \langle N_{\mathrm{part}}\rangle$\xspace}

\newcommand{\ncollav}{$\langle N_{\mathrm{coll}}\rangle$\xspace}

\newcommand{\pT}{$p_T$\xspace}
\newcommand{\highpT}{high-\pT}

\newcommand{\ridge}{ridge\xspace}
\newcommand{\pp}{$p$+$p$\xspace}
\newcommand{\Cu}{Cu+Cu\xspace}
\newcommand{\Au}{Au+Au\xspace}
\newcommand{\dAu}{$d$+Au\xspace}
\newcommand{\AplusA}{$A$+$A$\xspace}
\newcommand{\MeV}{MeV/$c$\xspace}
\newcommand{\GeV}{GeV/$c$\xspace}
\newcommand{\dphi}{$\Delta\phi$\xspace}
\newcommand{\deta}{$\Delta\eta$\xspace}
\newcommand{\vtwo}{$v_2$\xspace}
\newcommand{\vthree}{$V_{3\Delta}$\xspace}
\newcommand{\vthreesq}{$V_{3\Delta}$\xspace}
\newcommand{\vtwotildesq}{$V_{2\Delta}$\xspace}

\newcommand{\vthreeratio}{$V_{3\Delta}/V_{2\Delta}$\xspace}

\newcommand{\vthreeflowratio}{$(v_3/v_2)^2$\xspace}

\newcommand{\as}{away-side\xspace}
\newcommand{\ns}{near-side\xspace}
\newcommand{\nearside}{near-side\xspace}

\newcommand{\jl}{jet-like\xspace}
\newcommand{\PT}{PYTHIA\xspace}

\newcommand{\jetridgeeta}{$|\Delta\eta| < 0.78$\xspace}
\newcommand{\jetridgephi}{$|\Delta\phi| < 0.78$\xspace}
\newcommand{\ntrig}{$N_{\mathrm{trigger}}$\xspace}

\newcommand{\Fref}[1]{Figure~\ref{#1}}
\newcommand{\Tref}[1]{Table~\ref{#1}}
\newcommand{\Eref}[1]{Equation~\ref{#1}}

\newcommand{\roughly}{$\approx$\xspace}
\newcommand{\njet}{$Y_{J}$\xspace}
\newcommand{\nridge}{$Y_{\mathrm{ridge}}$\xspace}
\newcommand{\jlc}{jet-like correlation\xspace}
\newcommand{\jlcs}{jet-like correlations\xspace}
\newcommand{\jly}{jet-like yield\xspace}
\newcommand{\jlys}{jet-like yields\xspace}

\newcommand{\dsqnwitharg}{\frac{d^2N}{d\Delta\phi\, d\Delta\eta}(\Delta\phi,\Delta\eta)}

\newcommand{\dNdPhi}{\frac{dN}{d\Delta\phi}}
\newcommand{\dNdEta}{\frac{dN}{d\Delta\eta}}

\def\Deta{\mbox{$\Delta\eta$}}
\def\Dphi{\mbox{$\Delta\phi$}}
\newcommand{\detano}{\ensuremath{\Delta\eta}}
\newcommand{\dphino}{\ensuremath{\Delta\phi}}
\newcommand{\bEta}{b_{\Delta\eta}}

\newcommand{\Yjphi}{Y^{\Delta\phi}_{J}}
\newcommand{\Yjeta}{Y^{\Delta\eta}_{J}}

\def\Yridge{\mbox{$Y_{\mathrm{ridge}}$}}

\def\Argpseven{\mbox{$\left[-0.78,0.78\right]$}}
\def\Argaa{\mbox{$\left[-a,a\right]$}}

\begin{document}
  \linenumbers
\title{System size and energy dependence of \nearside di-hadron correlations}

\date{\today}

\input{authorList}

\begin{abstract}

Two-particle azimuthal (\dphi) and pseudorapidity (\deta) correlations using a trigger particle with large transverse momentum ($p_T$) in \dAu, \Cu and \Au collisions at \sNNsixtytwo and 200~GeV from the STAR experiment at RHIC are presented.    The \ns correlation is separated into a jet-like component, narrow in both \dphi and \deta, and the \ridge, narrow in \dphi but broad in \deta.  Both components are studied as a function of collision centrality, and the \jlc is studied as a function of the trigger and associated $p_T$. The behavior of the \jl component is remarkably consistent for different collision systems, suggesting it is produced by fragmentation.  The width of the \jlc is found to increase with the system size.  The \ridge, previously observed in \Au collisions at \sNNtwohundred, is also found in \Cu collisions and in collisions at \sNNsixtytwo, but is found to be substantially smaller at \sNNsixtytwo than at \sNNtwohundred for the same average number of participants (\npartav).  
Measurements of the \ridge are compared to models.

\end{abstract}


\maketitle
\section{Introduction}

Jets are a useful probe of the hot, dense medium created in heavy-ion collisions at the Relativistic Heavy Ion Collider (RHIC) at Brookhaven National Laboratory (BNL).
Jet quenching~\cite{Wang:1991xy} was first observed as the suppression of inclusive hadron spectra at large transverse momenta (\pT) in central \Au collisions with respect to \pp data scaled by number of binary nucleon-nucleon collisions ~\cite{Adler:2002xw,Adcox:2001jp,Adcox:2002pe,Adler:2003qi,Back:2003qr,Adams:2003kv}.  Properties of jets at RHIC have been studied extensively using di-hadron correlations relative to a trigger particle with large transverse momentum~\cite{starZYAM,STARConical:2008nd,DisappearanceAwaySide,PunchThrough,RidgePaper:2009qa,Abelev:2009jv,Agakishiev:2010ur}. 

Systematic studies of associated particle distributions on the opposite side of the trigger particle revealed their significant modification in \Au relative to \pp and \dAu collisions at the top RHIC energy of \sNNtwohundred.  For low \ptassoc, the amplitude of the \as peak is greater and the shape is modified in \Au collisions \cite{starZYAM, STARConical:2008nd}.  At intermediate \pT (\trigrange{4}{6}, \assocrangevar{2}), the \as correlation peak is strongly suppressed~\cite{DisappearanceAwaySide}. At higher \pT, the \as peak reappears without shape modification, but the \as per trigger yield is smaller in \Au collisions than in \pp and \dAu~\cite{PunchThrough}.

The associated particle distribution on the near side of the trigger particle, the subject of this paper, is also significantly modified in central \Au collisions.  In \pp and \dAu collisions, there is a peak narrow in azimuth (\dphi) and pseudorapidity (\deta) around the trigger particle, which we refer to as the \jlc.  This peak is also present in \Au collisons, but an additional structure which is narrow in azimuth but broad in pseudorapidity has been observed in central \Au collisions at \sNNtwohundred ~\cite{starZYAM,RidgePaper:2009qa,Abelev:2009jv,Agakishiev:2010ur}.  This structure, called the \ridge, is independent of \deta within the STAR acceptance, $|\Delta\eta|<$~2.0, within errors, and persists to high \pT (\pttrig \roughly 6 \GeV, \ptassoc \roughly 3 \GeV).  
While the spectrum of particles in the \jlc becomes flatter with increasing \pttrig, the slope of the spectrum of particles in the \ridge is independent of \pttrig and closer to the inclusive spectrum than to that of the \jlc.
Recent studies of di-hadron correlations at lower transverse momenta (\pttrig$>$~2.5~GeV/c, \ptassoc$>$~20~MeV/c) by the PHOBOS experiment show that the \ridge is roughly independent of \deta and extends over four units in \deta ~\cite{Alver:2009id}.  A similar broad correlation in pseudorapidity is also evident in complementary studies of minijets using untriggered di-hadron correlations~\cite{MiniJets,MiniJetsNew}.

Several mechanisms for the production of the \ridge have been proposed since the first observation of this new phenomenon. In one model~\cite{Armesto:2004pt} the \ridge is proposed to be formed from gluon radiation emitted by a high-\pT parton propagating in the medium with strong longitudinal flow. 
The momentum-kick model proposes that the \ridge forms as a fast parton traveling through the medium loses energy through collisions with partons in the medium, causing those partons to be correlated in space with the fast parton~\cite{Wong:2008yh,Wong:2007mf,Wong:2007pz}.  Parton recombination has been also proposed as a mechanism for the production of the \ridge~\cite{Hwa:2008jt,Hwa:2008um,Chiu:2008ht}. Another model~\cite{Voloshin:2003ud,Pruneau:2007ua} suggests that the \ridge is not actually caused by a hard parton but is the product of radial flow and the surface biased emission of the trigger particle, causing an apparent correlation between particles from the bulk and high-\pT trigger particles.

Another class of models is based on the conversion of correlations in the initial state into momentum space through various flow effects. The model in~\cite{Majumder:2006wi} explains the \ridge as arising from the spontaneous formation of extended color fields in a longitudinally expanding medium due to the presence of plasma instabilities. Long-range pseudorapidity correlations formed in an initial state glasma combined with radial flow have been also discussed as a mechanism for the \ridge~\cite{Dumitru:2008wn,Glasma2,Gavin:2008ev}.
Recently, it has been suggested that triangular anisotropy in the initial collision geometry caused by event-by-event fluctuations can give rise to triangular flow, which leads to the ridge and contributes to the double peaked \as observed in heavy-ion collisions at RHIC~\cite{Mishra:2007tw,Sorensen:2010zq,Takahashi:2009na,Alver:2010gr,Alver:2010grErratum,Alver:2010dn,Holopainen:2010gz}.

In this paper we present measurements of the system size and collision energy dependence of near-side di-hadron correlations using data from \Cu and \Au collisions at \sNNsixtytwo and \sNNtwohundred measured by the STAR experiment at RHIC.
In particular, we investigate the centrality dependence of the \jlcs and the \ridge and the transverse momentum dependence of the \jlcs. The properties of \jlcs in heavy-ion collisions are compared to those from \dAu collisions and \PT simulations to look for possible medium modifications and broadening of the near-side \jlc, for example, due to gluon bremsstrahlung~\cite{Armesto:2004pt}. 
The new results on the system size and energy dependence of the \ridge yield  presented in this article extend our knowledge of this phenomenon and, in combination with other measurements at RHIC, provide important quantitative input and constraints to model calculations.

\section{Experimental setup and data sample}

The results presented in this paper are based on data measured by the STAR experiment from \dAu collisions  at \sNNtwohundred in 2003, \Au collisions  at  \sNNsixtytwo and \twohundred in 2004, and \Cu collisions at \sNNsixtytwo and  \twohundred in 2005.  The \dAu events were selected using a minimally biased (MB) trigger requiring at least one beam-rapidity neutron in the Zero Degree Calorimeter (ZDC), located at 18~m from the nominal interaction point in the Au beam direction, accepting 95$\pm$3\% of the \Au hadronic cross section \cite{Adams:2003im}.  For \Cu collisions, the MB trigger was based on the combined signals from the Beam-Beam Counters (BBC) at forward rapidity (3.3~$<|\eta|<$~5.0) and a coincidence between the ZDCs. The MB trigger for \Au collisions at \sNNsixtytwo and \twohundred was obtained using a ZDC coincidence, a signal in both BBCs and a minimum charged particle multiplicity in an array of scintillator slats arranged in a barrel, the Central Trigger Barrel (CTB), to reject non-hadronic interactions.

\begin{table}[b!]
\caption{Number of events after cuts (see text) in the data samples analyzed.}
\label{tabevents}
\begin{tabular}{c|c|c|c}
System  & Centrality & $\sqrt{s_{{NN}}}$ [GeV] &  No. of events [10$^6$]\\ \hline
\Cu   &  0-60\% & 62.4   &  24\\ 
\Au   &  0-80\% & 62.4   &  8\\ 
\dAu    &  0-95\% & 200  &  3\\ 
\Cu   &  0-60\% & 200  &  38\\ 
\Au   &  0-80\% & 200  &  28\\ 
\Au   &  0-12\% & 200  &  17\\ 
\end{tabular}
\end{table}


For \Au collisions at \sNNtwohundred, an additional online trigger for central collisions was used. 
This trigger was based on the energy deposited in the ZDCs in combination with the multiplicity in the CTB. 
The central trigger sampled the most central 12\% of the total hadronic cross section.

In order to achieve a more uniform detector acceptance, only those events with the primary vertex position along the longitudinal beam direction ($z$) within 30~cm
of the center of the STAR detector were used for the analysis.  For the \dAu collisions this was expanded to $|z|~<$~50~cm.  
The number of events after the vertex cut in individual data samples is summarized in Table~\ref{tabevents}.

The STAR Time Projection Chamber (TPC)~\cite{STARNIM} was used for tracking of charged particles. The collision centrality was determined from the uncorrected number of charged  tracks at mid-rapidity ($|\eta|<$~0.5) in the TPC. The charged tracks used for the centrality determination had a three dimensional distance of closest approach (DCA) to the primary vertex of less than 3~cm and at least 10 fit points from the TPC. Each data sample was then divided into several centrality bins, and the fraction of the geometric cross section, the average number of participating nucleons (\npartav) and the average number of binary collisions (\ncollav) were calculated using Glauber Monte-Carlo model calculations~\cite{Miller:2007ri}.


\section{Data Analysis}

\subsection{Correlation technique}

Tracks used in this analysis were required to have at least 15 fit points in the TPC,
a DCA to the primary vertex of less than 1 cm, and a pseudorapidity $|\eta|<$~1.0.  As in previous di-hadron correlation studies in the STAR experiment~\cite{RidgePaper:2009qa,Catu:2009gg,MarkHornerPaper}, a \highpT trigger particle was selected and the raw distribution of associated tracks relative to that trigger in pseudorapidity (\deta) and azimuth (\dphi) is studied.  This distribution, ${d^2N_{\mathrm{raw}}}/{d\Delta\phi\, d\Delta\eta}$, is normalized by the number of trigger particles, \ntrig, and corrected for the efficiency and acceptance of associated tracks $\varepsilon$:
\begin{eqnarray}
\dsqnwitharg & = & \frac{1}{N_{\mathrm{trigger}}} \frac{d^2N_{\mathrm{raw}}}{d\Delta\phi d\Delta\eta} \nonumber \\
& & \frac{1}{\varepsilon_{\mathrm{assoc}}(\phi,\eta)} \frac{1}{\varepsilon_{\mathrm{pair}}(\Delta\phi,\Delta\eta)}.
\label{eq:di-hadronDist}
\end{eqnarray}
The efficiency correction $\varepsilon_{\mathrm{assoc}}(\phi,\eta)$ is a correction for the TPC single charged track reconstruction efficiency and $\varepsilon_{\mathrm{pair}}(\Delta\phi,\Delta\eta)$ is a correction for track merging and finite TPC track-pair acceptance in \dphi and \deta as described in detail below. 
The data presented in this paper are averaged between positive and negative \dphi and \deta regions and are reflected about \dphi=0 and \deta=0 in the plots. 

\subsection{Single charged track efficiency correction}
\begin{figure}[t!]
\rotatebox{0}{\resizebox{8.6cm}{!}{
\includegraphics{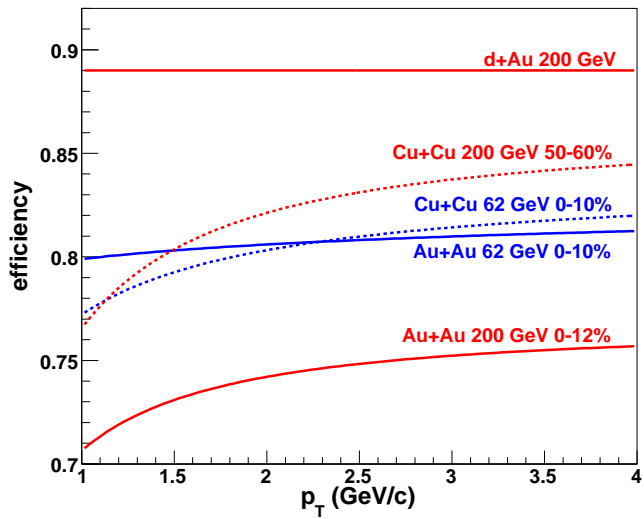}
}}
\caption{(Color online) Parameterizations of the transverse momentum dependence of the reconstruction efficiency of charged particles in the TPC in various collision systems, energies and centrality bins for the track selection cuts used in this analysis.  Note the zero suppression of the axes.}
\label{effcurves}
\end{figure}

The single charged track reconstruction efficiency in the TPC is determined by simulating the detector response to a charged particle and embedding these signals into a real event.
This hybrid event is analyzed using the same software as for the real events. The efficiency for detecting a single track as a function of \pT, $\eta$, and centrality is determined from the number of simulated particles which were successfully reconstructed.  The single track efficiency is approximately constant for \pT $>$~2~\GeV and ranges from around 75\% for central \Au events to around 85\% for peripheral \Cu events as shown in Figure~\ref{effcurves}.  The efficiency for reconstructing a track in \dAu is 89\%.  The systematic uncertainty on the efficiency correction, 5\%, is strongly correlated across centralities and \pT bins for each data set but not between data sets.  In the correlations, each track pair is corrected for the efficiency for reconstructing the associated particle.  Since the correlations are normalized by the number of trigger particles, no correction for the efficiency of the trigger particle is necessary.

\begin{figure*}[t!]
\includegraphics[width=15cm]{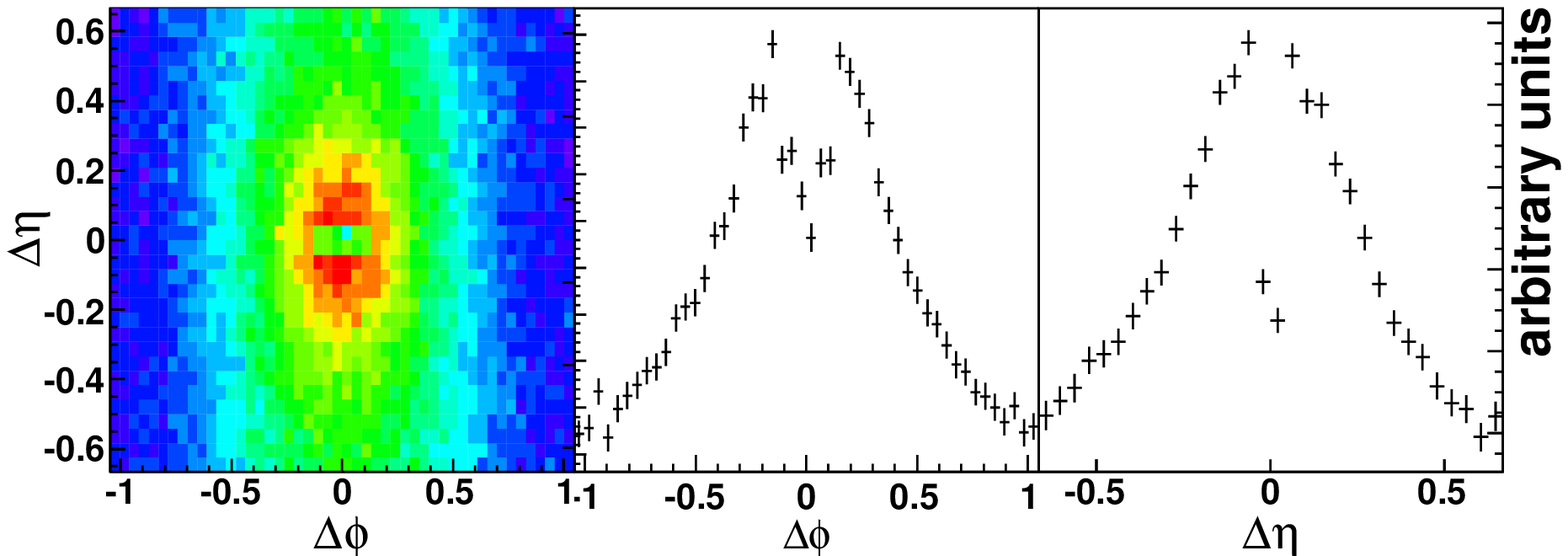} 
\caption{(Color online) Two-particle di-hadron correlation function in (\dphi, \deta) for \stdtrig and \stdassoc in 0-12\% central \Au collisions at \sNNtwohundred, not corrected for track merging (left). The projection of the correlation function in \dphi for $|\Delta\eta|<$~0.042 is shown in the middle and the projection in \deta for $|\Delta\phi|<$~0.17 on the right.}
\label{AuAu2D}
\end{figure*}

\subsection{Corrections for track merging and track crossing effects in the TPC}
\label{section-merging}

The reconstruction of charged tracks from TPC hits is performed iteratively, with hits removed from the event once they are assigned to a track. If two tracks have small angular separation in both pseudorapidity and azimuth, they are more difficult to reconstruct because distinct hits from each particle may not be resolved by the TPC. 
If two particles are close in momenta or have sufficiently high \pT that their tracks are nearly straight, they may not be distinguished.
This effect, called track merging, reduces the number of pairs observed at small opening angles and results in an artificial dip in the raw correlations centered at ($\Delta\phi,\Delta\eta$)~=~(0,0). 

Figure~\ref{AuAu2D} shows an example of a ($\Delta\phi$, $\Delta\eta$) two-particle di-hadron correlation function in central \Au collisions at $\sqrt{s_{NN}}$~=~200~GeV, along with the corresponding $\Delta\phi$ and $\Delta\eta$ projections  for \stdtrig and \stdassoc. For small angular separations, a clear dip in the raw correlations is visible and must be corrected for in order to extract the yield of associated particles on the near side.

\begin{figure}[b!]
\includegraphics[width=8.5cm]{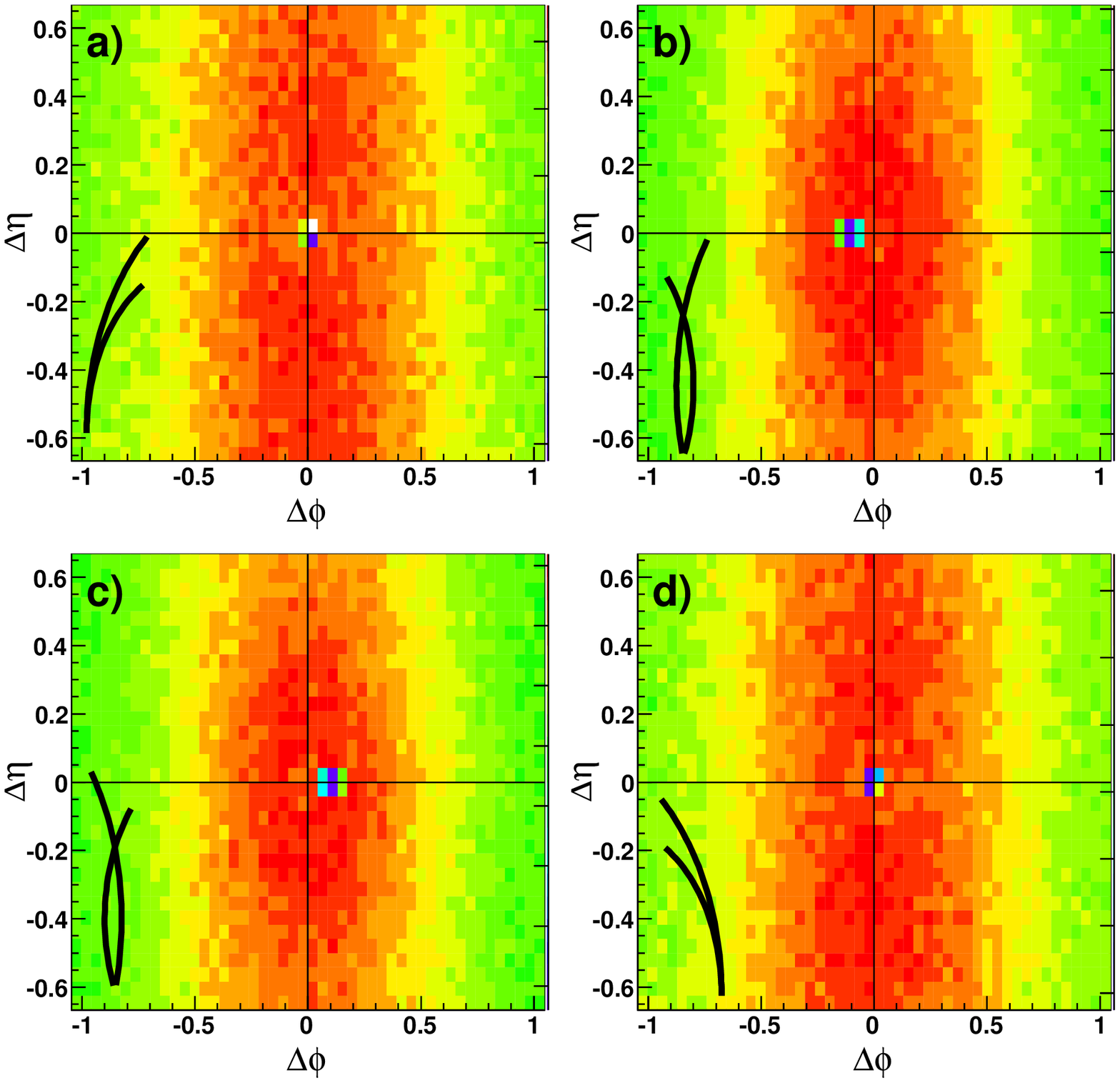} 
\caption{(Color online) The dip region in (\dphi, \deta) uncorrected di-hadron correlations in 0-12\% central \Au collisions at \sNNtwohundred for \stdtrig and \stdassoc in the four helicity combinations of trigger and associated particles: (a) ($h_{\mathrm{trig}},h_{\mathrm{assoc}}$)~=~(1,1),
(b) ($h_{\mathrm{trig}},h_{\mathrm{assoc}}$)~=~(1,-1), (c) ($h_{\mathrm{trig}},h_{\mathrm{assoc}}$)~=~(-1,1), and (d) ($h_{\mathrm{trig}},h_{\mathrm{assoc}}$)~=~(-1,-1).  Cartoons indicate which dips are from track merging and which are from track crossing.}
\label{4dips}
\end{figure}

Another similar effect is evident in the active TPC volume from high-\pT tracks which cross. While few hits are lost in this case, one track may lose hits near the crossing point and therefore be split into two shorter tracks. Shorter tracks are less likely to meet the track selection criteria. 
Track merging and track crossing cause four dips near ($\Delta\phi,\Delta\eta$)~=~(0,0) but slightly displaced in $\Delta\phi$. The location and width of these dips in $\Delta\phi$ is dependent on the relative helicities, $h$, and the $p_T$ intervals of the trigger and associated particles. The helicity $h$ is given by

\begin{equation}
h = \frac{-qB}{|qB|}
\end{equation}
\noindent where $q$ is the charge of the particle and $B$ is the magnetic field.  The dips for tracks of the same helicity are dominantly due to track merging and the dips for tracks of opposite helicities are dominantly due to track crossing.
Figure~\ref{4dips} displays the correlation function from Figure~\ref{AuAu2D} in four different helicity combinations of trigger and associated particles showing the finer substructure of the dip on the near side.
When the helicities of the trigger and associated particles are the same, the percentage
of overlapping hits is greater.    
Because higher \pT tracks have a smaller curvature, it is more likely for two high \pT tracks close in azimuth and pseudorapidity to be merged than lower \pT particles.
However, track pairs are lost whether the pair is part of the combinatorial background or part of the signal.  This effect means that the magnitude of the dip is greater in central collisions where the background is greater and decreases with increasing $p_T$ because of the decreasing background.

\begin{figure}[b!]
\includegraphics[width=8.5cm]{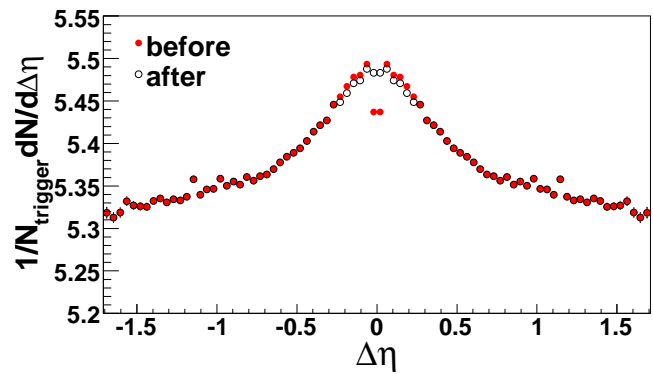} 
\caption{(Color online) The raw correlation in \deta for di-hadron correlations for \stdtrig and \stdassoc for 0-12\% central \Au collisions for $|\Delta\phi|<$~0.78 before and after the track merging correction is applied.  The data have been reflected about \deta=0.}
\label{beforeafter}
\end{figure}

One way to correct for the track merging effect is to remove pairs from mixed events that would have merged in real data. 
The environment in mixed events must be similar to real data in order for pair rejection to be accurately 
reproduced. The reference multiplicities  of both events were required to be within 10 of each other to assure similar track density. In addition, mixed events were required to have vertices within 2 cm of each other along
the beam axis in order to ensure similar geometric acceptance to avoid a different dip shape in $\Delta\eta$. In order to calculate accurately the percentage of merged hits, the origin of the associated track was shifted to the vertex of the event which the trigger particle originated from.

Previous analyses of low  momentum tracks have shown that eliminating pairs from both data and mixed 
events with a fraction of merged hits greater than 10\% was sufficient to correct 
for merging~\cite{STARHBT}. 
By discarding pairs with more than 10\% shared hits we insure that  the percentage of merged track pairs is the same in the data and the mixed events.

 The correlation function for a given helicity combination of trigger and associated particles was corrected by mixed events.  After this correction, a small residual dip
remains, mostly due to track crossing.  While the mixed events correct for the dip due to true track merging well, they do not correct for track crossing as well. The remaining dip is then corrected for using the symmetry of the correlations. Since the data should be symmetric
about $\Delta\phi$~=~0, the data on the same side as the dip are discarded and replaced by
the data on the side without the dip. Then the data are reflected about $\Delta\eta$~=~0 and
added to the unreflected data to minimize statistical fluctuations.
This method is only applied to $|\Delta\phi|<$~1.05 and $|\Delta\eta|<$~0.67, the region shown in \Fref{AuAu2D}, because it is computationally intensive and track merging and track crossing only affect small angular separations. For large $\Delta\phi$ and $\Delta\eta$, the method described in the next section is applied. The track merging correction is done for each di-hadron correlation function separately with the appropriate cuts on \pttrig, \ptassoc and collision centrality. An example of the di-hadron correlation function before and after the track merging correction is shown in Figure~\ref{beforeafter}.  The slight decrease in the correlation function for some data points is an artifact of the correction procedure and reflects the uncertainty in the correction.

\subsection{Pair acceptance correction}\label{acceptanceCorrection}
With the restriction that each track falls within $|\eta|<1.0$, there is a limited acceptance for track pairs.  For \deta \roughly~$0$, the geometric acceptance of the TPC for track pairs is \roughly 100\%, however, near \deta \roughly~$2$ the acceptance is close to 0\%.  In azimuth the acceptance is limited by the 12 TPC sector boundaries, leading to dips in the acceptance of track pairs in azimuth.  To correct for the geometric acceptance, the distribution of tracks as a function of $\eta$ and $\phi$ was recorded for both trigger and associated particles.  A random $\eta$ and $\phi$ was chosen from each of these distributions to reconstruct a random \deta and \dphi for each selection of \pttrig, \ptassoc, and centrality.  This was done for at least four times as many track pairs as in the data and was used to calculate the geometrical acceptance correction for pairs.

\subsection{Subtraction of anisotropic elliptic flow background}\label{vtwosection}

Correlations of particles with the event plane due to anisotropic flow (\vtwo) in heavy-ion collisions are indirectly reflected in di-hadron correlations and have to be subtracted for studies of the \ridge. This background in \dphi over the interval \Argaa\xspace is approximated by

\begin{align}
\label{eq:BDphi}
B_{\Delta\phi} \Argaa \; & \equiv \; \notag
 \\ b_{\Delta\phi} & \int_{-a}^{a}d\Dphi
\left(1+2 \langle v_{2}^{\mathrm{trig}}\rangle \langle v_{2}^{\mathrm{assoc}} \rangle \cos2\Dphi\right) 
\end{align}\label{Eq:bdef}

\noindent where $a$ is chosen to be 0.78 so that the majority of the signal is contained~\cite{RidgePaper:2009qa,Catu:2009gg}.
The level $b_{\Delta\phi}$ of the background is determined using the Zero Yield At Minimum (ZYAM) method~\cite{starZYAM}. The level of the background is taken as the value of the minimum bin.  The systematic errors due to the choice of the minimum bin rather than either of the two neighboring bins are negligible compared to the systematic errors due to the magnitude of \vtwo, discussed below.

The ZYAM method is commonly used for di-hadron correlations at RHIC, for example ~\cite{PunchThrough,starZYAM,Adare:2008cqb}, and is justified if the near- and away-side peaks are separated by a 'signal-free' region.  At lower transverse momenta ($p_T<$~2 GeV/c) and in central \Au collisions at \sNNtwohundred the broadening of the away-side correlation peak may cause overlap of the near- and away-side peaks  and consequently makes the ZYAM normalization procedure biased. Alternatively, a decomposition of the correlation function using a fit function containing the anisotropic flow modulation of the combinatorial background and components describing the shape of the correlation peaks could be used as well~\cite{Adams:2006tj,Trainor:2009gj}.  In this paper we use the ZYAM prescription to remain consistent with our earlier measurements of the near-side ridge~\cite{RidgePaper:2009qa}.  ZYAM used with conservative bounds on \vtwo will, if anything, underestimate the \ridge yield.

For all collision systems and energies studied, the uncertainty bounds on \vtwo were determined by comparing different methods for the \vtwo measurement.  We assume that the error on the \vtwo of the trigger and associated particles is 100\% correlated.  Event plane measurements of flow and two-particle measurements such as the two-particle cumulant method are sensitive to non-flow from sources such as jets.  These methods may overestimate \vtwo.  Methods such as the four-particle cumulant method are less sensitive to contributions from jets, however, these methods may over-subtract contributions from event-by-event fluctuations in \vtwo.  Therefore these methods underestimate the \vtwo that should be used for the background subtraction in di-hadron correlations~\cite{Voloshin:2008dg}.  For each system at least one measurement which may overestimate \vtwo and at least one measurement which may underestimate \vtwo is included.  \vtwo and systematic errors on \vtwo for \Au collisions at \sNNsixtytwo are from comparisons of the event plane method using the forward TPCs for the event plane determination and the four particle cumulant method~\cite{AuAu62Flow}.  The \vtwo and systematic errors on \vtwo for \Au collisions at \sNNtwohundred are as described in~\cite{RidgePaper:2009qa}.  The upper bound on \vtwo is from the event plane method using the forward TPCs for the determination of the event plane, the lower bound comes from the four particle cumulant method, and the average of the two is the nominal value.  \vtwo for \Cu collisions at \sNNsixtytwo and 200 GeV is from~\cite{CuCuFlow:2010tr}.  The nominal value is given by \vtwo from the event plane method using the forward TPCs for the determination of the event plane and the upper bound is from the statistical error for both \sNNsixtytwo and \twohundred.  For \Cu collisions measurements using the four particle cumulant method were not possible due to limited statistics.  Instead, for \Cu collisions at \sNNtwohundred the lower bound is determined by 
the magnitude of the cos($2\Delta\phi$) term extracted from fits to \pp data,
scaling it by \ncollav, and subtracting it from \vtwo determined using the event plane method to estimate the maximum contribution from nonflow in \Cu collisions.  For \Cu collisions at \sNNsixtytwo the systematic error is assumed to be the same in \Cu collisions at both energies.

With both methods for subtracting the \ridge contribution to the \jly described below, the systematic errors due to \vtwo cancel out in the \jly, assuming that \vtwo is independent of \deta in the TPC acceptance.  This assumption is  based on the measurements of \vtwo as a function of $\eta$~\cite{phobosFlow1,phobosFlow2}.

\subsection{Yield extraction}
\label{yieldmethod}
To quantify the strength of the \ns correlation it is assumed that it can be decomposed into a jet-like component, narrow in both azimuth and pseudorapidity, and a \ridge component which is independent of \deta. This approach is consistent with the method in~\cite{RidgePaper:2009qa}.  For the kinematic cuts applied to \pttrig and \ptassoc, the \jlc is contained within the cuts used in this analysis, \jetridgeeta and \jetridgephi.

To study the \jlc and the \ridge quantitatively we adopt the notation from ~\cite{RidgePaper:2009qa}, and introduce the projection of the di-hadron correlation function from Eq.~(\ref{eq:di-hadronDist}) onto the \deta axis:

\begin{eqnarray}
\label{eq:IDphi}
\left.\dNdEta\right|_{a,b}
\equiv\int_{a}^{b}d\Dphi\frac{d^2N}{d\Dphi{d}\Deta};
\end{eqnarray}

\noindent
and similarly on the \dphi axis:

\begin{eqnarray}
\label{eq:IDeta}
\left.\dNdPhi\right|_{a,b}
\equiv
\int_{|\detano| \in [a,b]}d\Deta\frac{d^2N}{d\Dphi{d}\Deta}
.
\end{eqnarray}
\noindent


To determine the \jl yield of associated charged particles two methods are used.  The first method is based on $\Delta\phi$ projections. Under the assumption that the \jly is confined within $|\Delta\eta|<$~0.78, subtracting the \dphi projections:
\begin{eqnarray}
\label{eq:NJphi}
\frac{dN_{J}}{d\Dphi}\left(\Dphi\right)=
\left.\dNdPhi\right|_{0,0.78}-\frac{0.78}{1.0}\left.\dNdPhi\right|_{0.78,1.78}
\end{eqnarray}
\noindent
removes both the elliptic flow and ridge contributions. Since the second \dphi projection is calculated in a larger \deta window, it has to be scaled by a factor 0.78/1.0, the ratio of the \deta width in the region containing the \jlc, the \ridge, and the background to the width of the region containing only the \ridge and the background.  This subtracts both the \ridge and \vtwo simultaneously since within errors both are independent of \deta ~\cite{phobosFlow1,phobosFlow2,RidgePaper:2009qa,Alver:2009id}.

\begin{figure}[t!]
\rotatebox{0}{\resizebox{9.0cm}{!}{
	\includegraphics{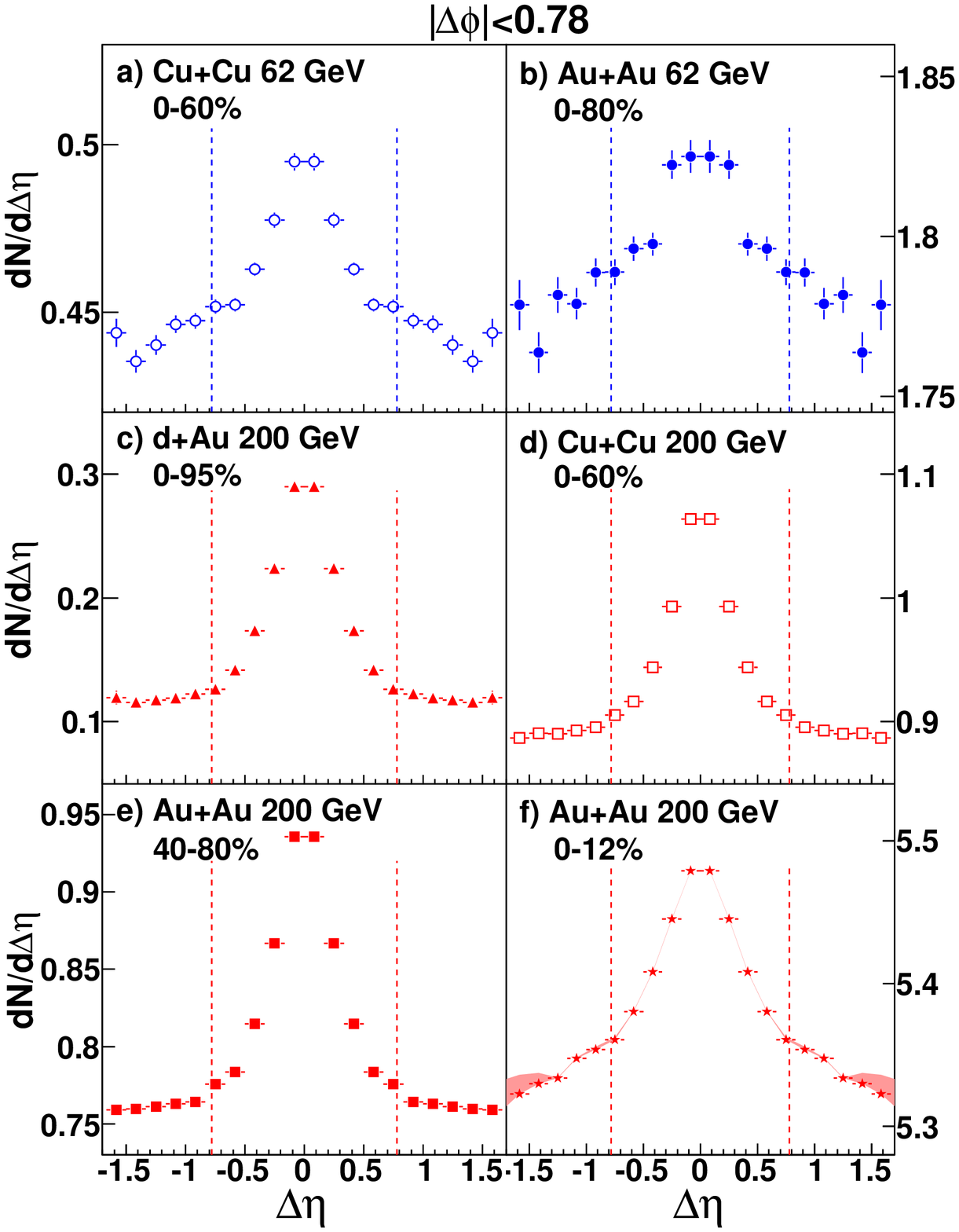}
}}
\vspace*{-0.5cm}
\caption{Color online) Sample correlations in \deta ($|\Delta\phi|<$~0.78) for \stdtrig and \stdassoc for (a) 0-60\% \Cu at \sNNsixtytwo (b) 0-80\% \Au at \sNNsixtytwo, (c) 0-95\% \dAu at \sNNtwohundred,  (d) 0-60\% \Cu at \sNNtwohundred, (e) 40-80\% \Au at \sNNtwohundred, and (f) 0-12\% central \Au at \sNNtwohundred.  Lines show the \deta range where the \jly is determined.  The data are averaged between positive and negative \deta and reflected in the plot.  Shaded lines in (f) show the systematic errors discussed in ~\ref{yieldmethod}.}\label{Figure1a}
\end{figure}

\begin{figure}[t!]
\rotatebox{0}{\resizebox{9.0cm}{!}{
	\includegraphics{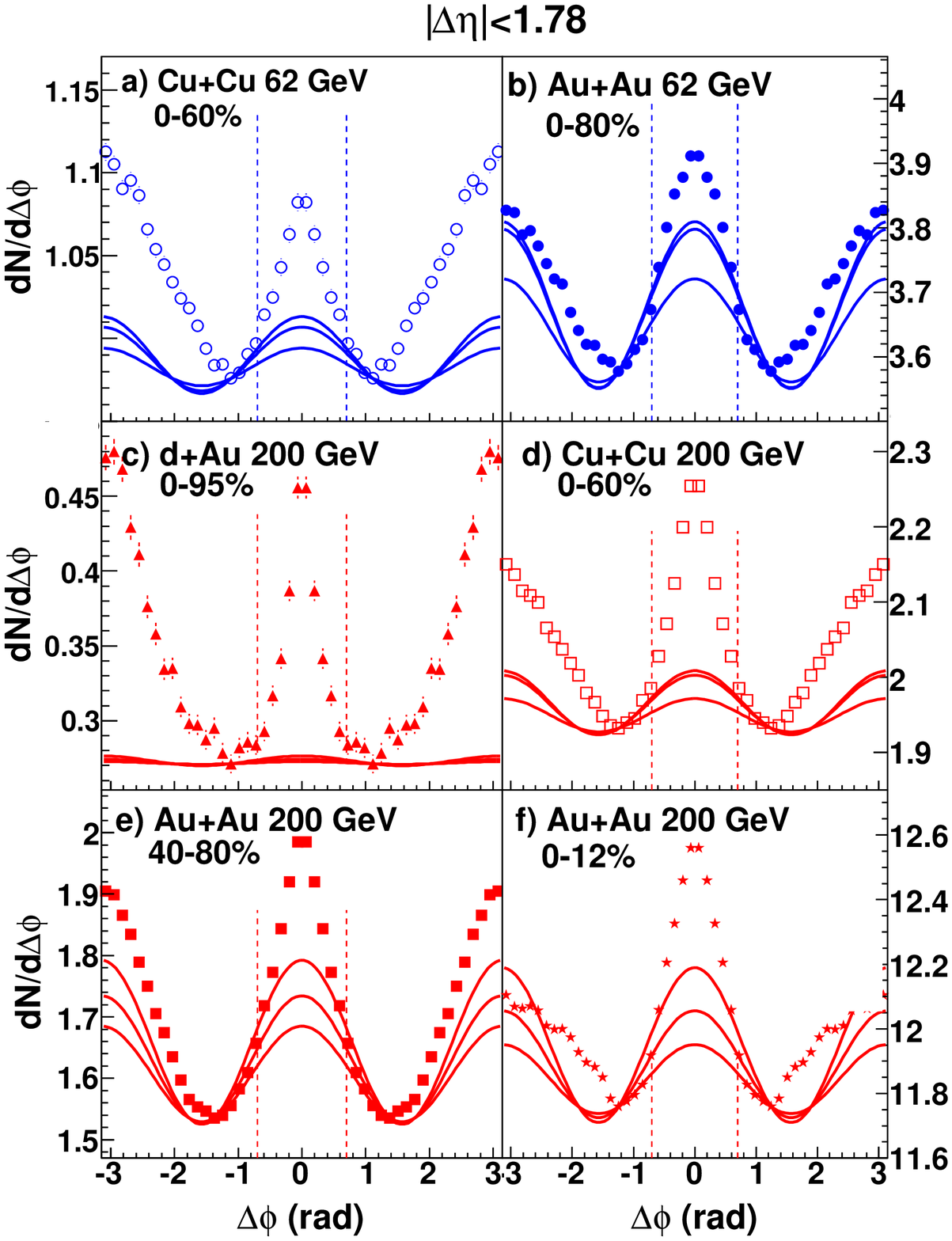}
}}
\vspace*{-0.5cm}
\caption{(Color online) Sample correlations in \dphi ($|\Delta\eta|<$~1.78) for \stdtrig and \stdassoc for (a) 0-60\% \Cu at \sNNsixtytwo (b) 0-80\% \Au at \sNNsixtytwo, (c) 0-95\% \dAu at \sNNtwohundred,  (d) 0-60\% \Cu at \sNNtwohundred, (e) 40-80\% \Au at \sNNtwohundred, and (f) 0-12\% central \Au at \sNNtwohundred. Solid lines show the estimated background using the ZYAM method with the range of \vtwo used for the determination of the systematic errors.  The data are averaged between positive and negative \dphi and reflected in the plot.   Vertical dashed lines show the \dphi range where the \jlc is determined.}\label{Figure1b}
\end{figure} 

\noindent The \jly $\Yjphi$ is then obtained by integrating Eq.(\ref{eq:NJphi})
\begin{align}
\Yjphi =  \int_{-0.78}^{0.78} d\dphino \; &
\frac{dN_{J}}{d\Dphi}\left(\Dphi\right)\,. \label{eq:Yjetphi} 
\end{align}
The second method for \jly determination is based on the \deta projection at the \ns: 
\begin{eqnarray}
\label{eq:NJeta}
\frac{dN_{J}}{d\Deta}\left(\Deta\right)=
\left.\dNdEta\right|_{-0.78,0.78}-b_{\Delta\eta}
\end{eqnarray}
\noindent
as \vtwo is  independent of pseudo-rapidity within the STAR acceptance and therefore only leads to a constant offset included in $\bEta$. The background level  $\bEta$ is determined by fitting a constant background $b_{\Delta\eta}$ plus a Gaussian to $\frac{dN_{J}}{d\Deta}\left(\Deta\right)$.
The yield determined from fit is discarded to avoid any assumptions about the shape of the peak and instead we integrate
Eq.~(\ref{eq:NJeta}) over \deta using bin counting to determine the \jly $\Yjeta$:

\begin{align}
\Yjeta =  \int_{-0.78}^{0.78} d\detano \; &
\frac{dN_{J}}{d\Deta}\left(\Deta\right)\,. \label{eq:Yjeta} 
\end{align}
\noindent  

The ridge yield \nridge is determined by first evaluating Eq.~(\ref{eq:IDeta}) over the entire \deta region to get $\frac{dN}{d\Dphi}$ then subtracting the modulated elliptic flow background $B_{\Delta\phi}$ and the jet-like contribution $\Yjeta$:
\begin{align}
\Yridge =  \int_{-0.78}^{0.78} d\dphino \; &
\left.\dNdPhi\right|_{0,1.78} \notag \\
& -  B_{\Delta\phi}\Argpseven - \Yjeta. \label{eq:Yridge}
\end{align}

\noindent  We determined the systematic error on the \njet due to uncertainty in the acceptance correction by comparing the mixed event method described in ~\ref{acceptanceCorrection} to the standard event mixing method and to a sample with a restricted $z$ vertex position.  The largest difference was seen in the central \Au data at \sNNtwohundred. To be conservative this difference is used as the systematic error for all the data sets.  The resulting systematic errors are listed in \Tref{errortable}.
This error is also present for the \ridge, since the \ridge is determined by subtracting \njet.  
Additionally, the systematic error on \njet due to the track merging correction does not exceed 1\%, the maximum size of the correction in the kinematic region studied in this paper.  This correction does not impact \nridge.

\begin{table}[b!]
\caption{Systematic uncertainties in the acceptance correction.}
\label{tabsystyield}
\begin{tabular}{c|c|c|c}
\pttrig &  \ptassoc & sys.error  & sys. error \\
 (GeV/$c$) &  (GeV/$c$)  & yield & Gaussian width \\ \hline
2.0-2.5 & $>$1.5 & $<$27\% & $<$10\%\\ 
2.5-3.0 & $>$1.5 & $<$18\% & $<$6\%\\
3.0-6.0 & $>$1.0 & $<$16\% & $<$6\% \\
3.0-6.0 & $>$1.5 & $<$9\% &  $<$6\% \\
\end{tabular}\label{errortable}
\end{table}

\subsection{2D Fits}
 In addition to the standard ZYAM procedure, we also analyzed the distribution of particles in Eq.~(\ref{eq:di-hadronDist})  using two-dimensional fits of the form:

\begin{eqnarray}
 \frac{d^2N}{d\Delta\phi d\Delta\eta} = &A&(1+\sum_{n=1}^{4}2V_{n\Delta}\cos (n\Delta\phi))\notag \\ 
 &+& \frac{Y_{J}}{2\pi\sigma_{\Delta\phi,J}\sigma_{\Delta\eta,J}} e^{-\frac{\Delta\eta}{2\sigma_{\Delta\eta,J}^{2}}}e^{-\frac{\Delta\phi}{2\sigma_{\Delta\phi,J}^{2}}}.
\label{eq:2dfit} 
\end{eqnarray}

\noindent with first four coefficients $V_{1\Delta}$, $V_{2\Delta}$ , $V_{3\Delta}$ and $V_{4\Delta}$ of a Fourier expansion and a term accounting for the \jlc on the \ns.  This approach is motivated by the class of models for \ridge production through a triangular initial condition.  If non-flow contributions are negligible, $V_{2\Delta}$ corresponds to the average of the product of the trigger particle \vtwo and the associated particle \vtwo and $V_{3\Delta}$ corresponds to the average of the product of the trigger particle $v_3$ and the associated particle $v_3$.  We use Eq.~(\ref{eq:2dfit})  to fit the data and extract \vthreeratio for all collision energies and systems. We allow $V_{3\Delta}$ to be negative.  A narrow roughly Gaussian \as peak at \dphi $\approx \pi$, which could arise from correlations from the production of an \as jet, would have a negative contribution to $V_{3\Delta}$ and a negative $V_{3\Delta}$ could indicate that flow is not the dominant production mechanism for these correlations.
Furthermore, $V_{2\Delta}$ is not constrained to the experimental values measured for \vtwo through other means.
There is no systematic error on \vthreeratio due to the efficiency because any uncertainty in the efficiency would change the magnitude of the modulations, given by A in Eq.~(\ref{eq:2dfit}), but not the relative size of those modulations, $V_{2\Delta}$ and $V_{3\Delta}$.  The uncertainty due to the fit and uncertainty in the acceptance correction is determined by fixing the parameters within the range given in \Tref{errortable}.  This gives an uncertainty of $<$4\% on \vthreeratio.

\begin{figure}[b!]
\rotatebox{0}{\resizebox{9cm}{!}{
\includegraphics{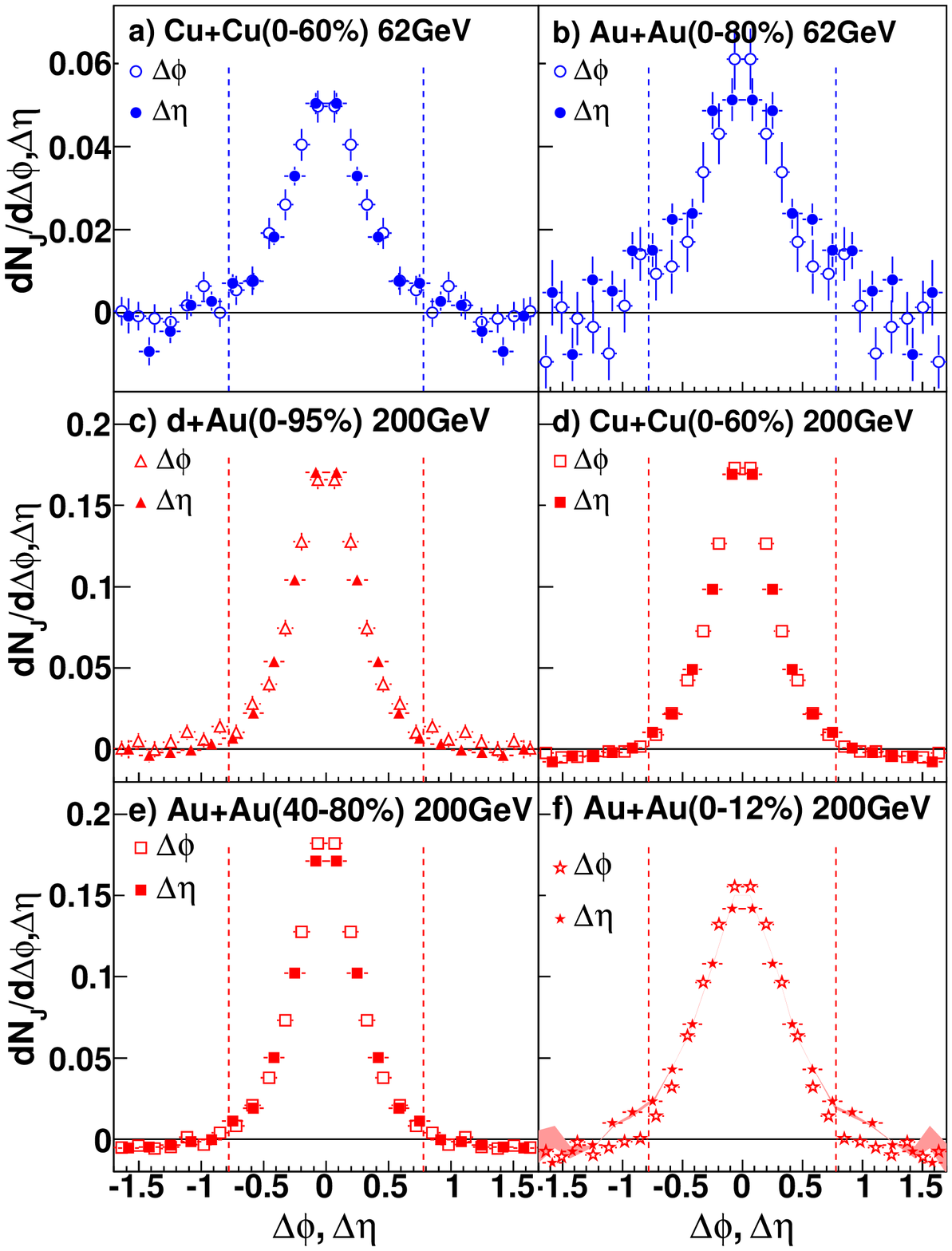}	
}}
\caption{(Color online) Background subtracted sample correlations for \stdtrig and \stdassoc on the near-side for  (a) 0-60\% \Cu at \sNNsixtytwo, (b) 0-80\% \Au at \sNNsixtytwo, (c) 0-95\% \dAu collisions at \sNNtwohundred, (d) 0-60\% \Cu at \sNNtwohundred, (e) 40-80\% \Au at \sNNtwohundred and (f) 0-12\% central \Au at \sNNtwohundred.  The dependence of the \jlc is shown as a function of \dphi ($|\Delta\eta|<$~1.78)  in open symbols and of \deta ($|\Delta\phi|<$~0.78) in closed symbols.  Lines show the \dphi  and \deta ranges where the \jly is determined.  The data are averaged between positive and negative \deta (\dphi) and reflected in the plot.    Lines in (f) show the systematic errors on $\frac{dN_{J}}{d\Deta}$ discussed in ~\ref{yieldmethod}.}\label{SampleCorrSubtr}
\end{figure}
\section{Results}

\subsection{Sample Correlations}

\Fref{Figure1a} shows fully corrected \deta projections of sample correlations on the near side ($|\Delta\phi|<$~0.78) before background subtraction for \dAu, \Cu and \Au collisions at energy \sNNsixtytwo and \sNNtwohundred.  The trigger particles were selected with transverse momentum \stdtrig and the associated particles with \stdassoc. The data show a clear jet-like peak sitting on top of the background and the ridge. The level of the background is increasing with energy and system size as expected, as more bulk particles are produced in the collision.

Examples of the complementary projections in \dphi  before background subtraction for ($|\Delta\eta|<$~1.78) are shown in \Fref{Figure1b}.  The elliptic flow modulated background is shown as solid curves. The middle curve corresponds to background calculated with the nominal value of \vtwo. The upper (lower) curve corresponds to the background if the upper (lower) bound on \vtwo is used instead.  Since we have conservatively assumed that the error on the \vtwo of the trigger and associated particles is 100\% correlated, the background occasionally goes above the signal in \Fref{Figure1b}(f) on the away-side.  However, since we are focusing the \ns we prefer this conservative estimate. Note that the uncertainty in the size of the elliptic flow modulated background affects only the ridge yield but not the \jly, since the elliptic flow contribution to the jet-like yield in \dphi cancels out in Eq.(\ref{eq:NJphi}) and in \deta is included in $\bEta$ in Eq.(\ref{eq:NJeta}).

Sample background subtracted correlation functions $\frac{dN_{J}}{d\Deta}$ from Eq.(\ref{eq:NJeta}) and $\frac{dN_{J}}{d\Dphi}$ from Eq.(\ref{eq:NJphi}) on the near side for \stdtrig and \stdassoc from \Fref{Figure1a} and  \Fref{Figure1b} are shown in \Fref{SampleCorrSubtr}.
For the given kinematic selection, the extracted \jlc peaks in both \deta and \dphi projections look very similar in  all studied systems and collision energies. The \jlys discussed through the rest of the paper are obtained from the \deta projection method; the \dphi method is only used for determining the width of the \jlc in \dphi. Below, the dependence of the near side \jly and Gaussian width of the \jlc peak on collision centrality and the transverse momentum of the trigger and associated particles are studied in detail.

\begin{figure}[t!]
\rotatebox{0}{\resizebox{8.6cm}{!}{
	\includegraphics{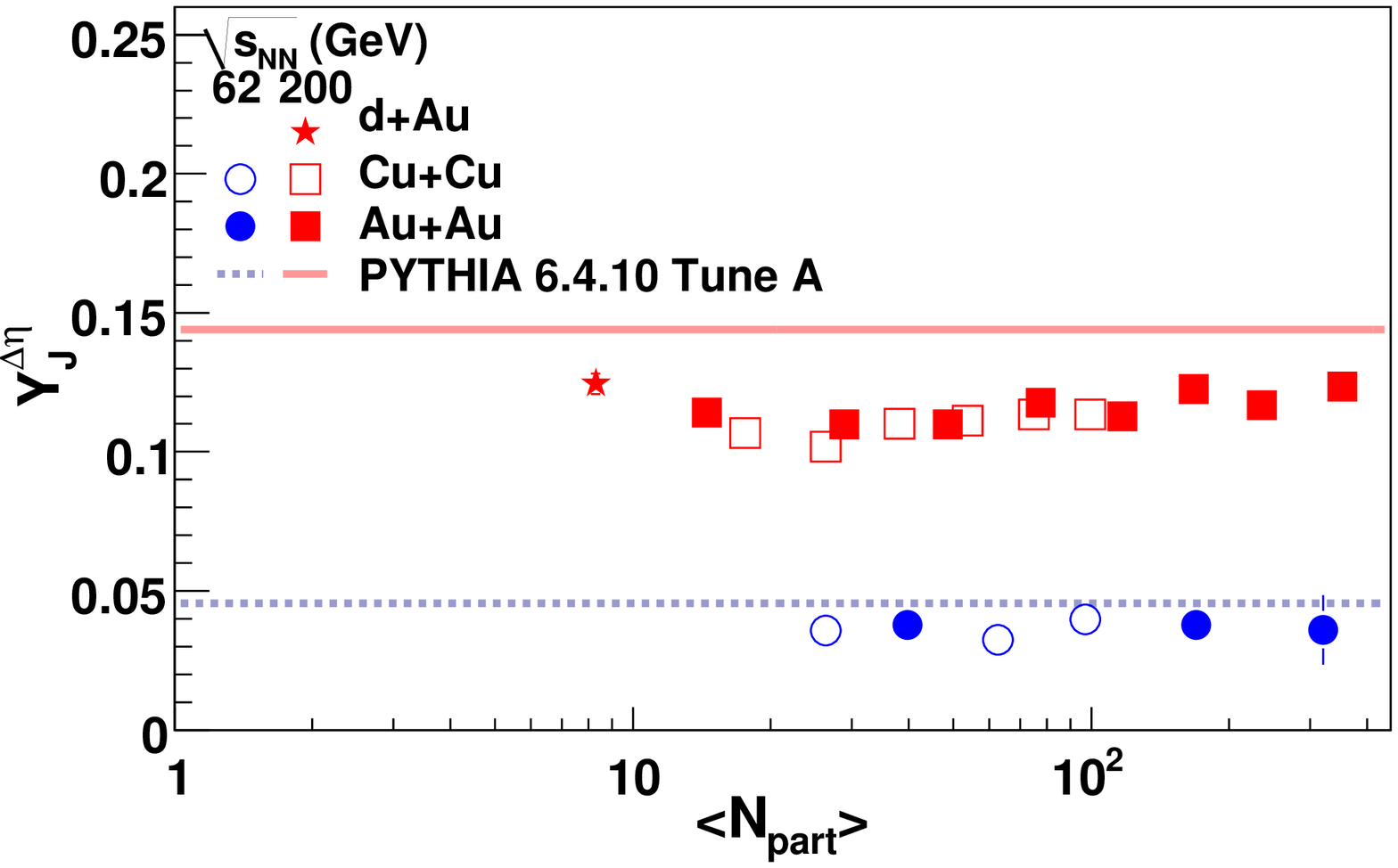}
}}
\caption{(Color online) Dependence of \jly on \npartav for \stdtrig and \stdassoc for \Cu and \Au at \sNNsixtytwo and \dAu, \Cu and \Au at \sNNtwohundred.  Comparisons to \PT, \npartav=2, are shown as lines.  The 5\% systematic error due to the uncertainty on the associated particle's efficiency is not shown and systematic errors due to the acceptance correction are given in \Tref{errortable}.  The background level and \vtwo values used for the extraction of these yields are given in \Tref{tabnpart}.}\label{Figure3}
\end{figure}

\begin{figure}[t!]
\rotatebox{0}{\resizebox{8.6cm}{!}{
	\includegraphics{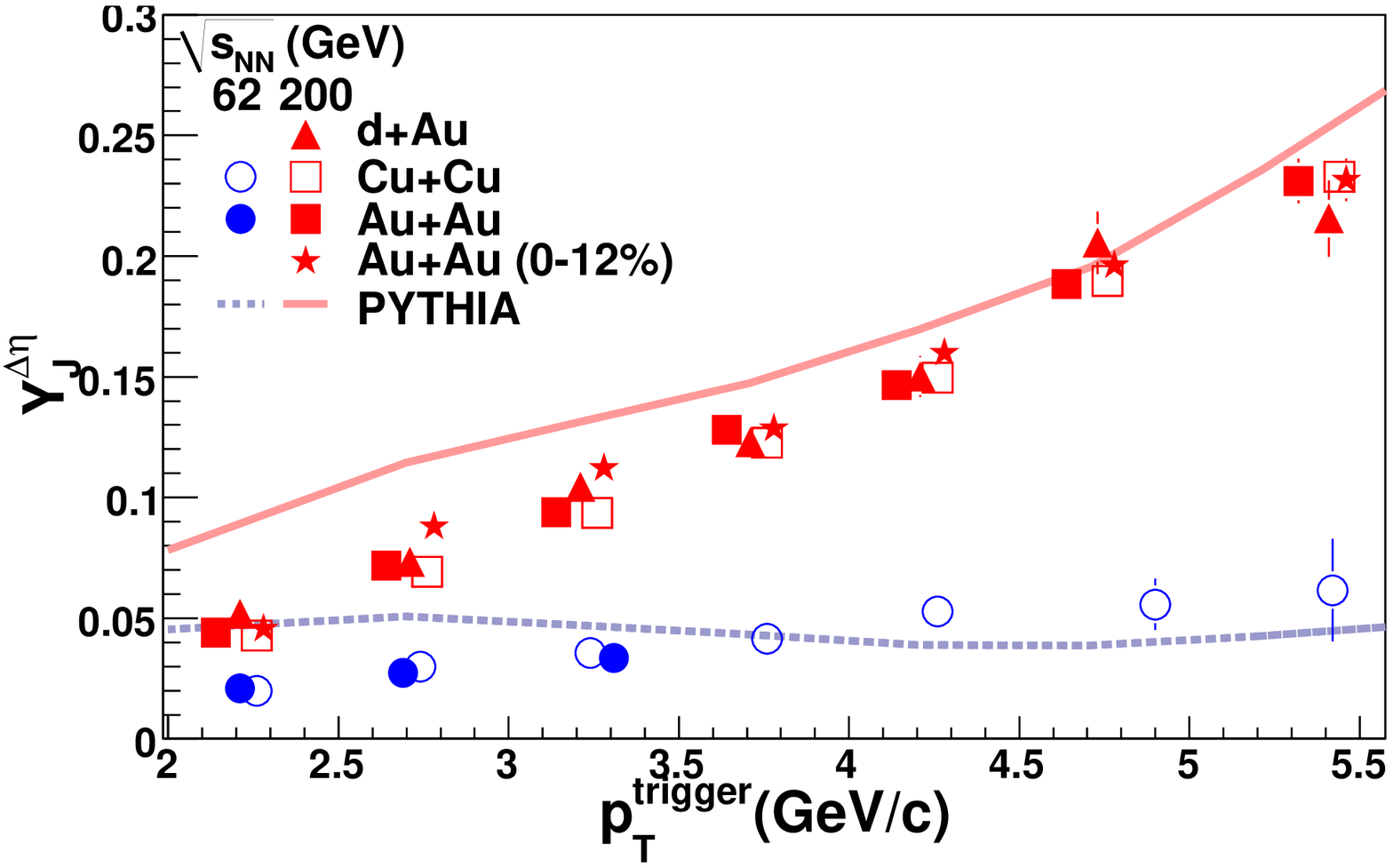}
}}
\caption{(Color online) Dependence of \jly on \pttrig for 0-95\% \dAu, 0-60\% \Cu at \sNNsixtytwo and \sNNtwohundred, 0-80\% \Au at \sNNsixtytwo, and 0-12\% and 40-80\% \Au at \sNNtwohundred.  Comparisons to \PT are shown as lines.  The 5\% systematic error due to the uncertainty on the associated particle's efficiency is not shown and systematic errors due to the acceptance correction are given in \Tref{errortable}.}\label{Figure1}
\end{figure} 

\subsection{The near-side jet-like component}

The centrality dependence of the \jly for \stdtrig and \stdassoc is plotted in \Fref{Figure3}.  The \jly at \sNNsixtytwo is lower than that at \sNNtwohundred by about a factor of three, which can be understood as the result of a steeply falling jet spectrum folded with the fragmentation function.
The measured yields are compared to \PT simulations shown as a line in \Fref{Figure3}. For these studies \PT version 6.4.10 \cite{Sjostrand:2006za} CDF tune A \cite{Field:2002vt}, which matches the data from the Tevatron at $\sqrt{s}$~=1.8~TeV and also describes the pion and proton inclusive spectra well at RHIC energies~\cite{Adams:2006xb,Adams:2006nd,Adams:2003qm}, is used. The \PT prediction is somewhat above the data, even for \dAu collisions.  However, considering the fact that the data are from heavy-ion collisions and the \PT simulations are for \pp collisions, good agreement is unanticipated.  For a given number of participating nucleons, \npartav, and collision energy, \sNN, there is no significant difference between the \dAu,  \Cu and \Au collisions observed, as expected if the \jlc were dominantly produced by vacuum fragmentation.

 The dependence of the \jly on \pttrig for \stdassoc is plotted in \Fref{Figure1} for all studied collision systems and energies.  Data from \Au collisions at \sNNtwohundred are shown separately for peripheral (40-80\%) and central (0-12\%) \Au collisions.  The \jly increases constantly with \pttrig for both \Cu and \Au and for \sNNsixtytwo and \sNNtwohundred. The effect of a steeper jet spectrum at \sNNsixtytwo relative to \twohundred discussed above is now reflected in the difference between the \jlys at the two energies. This difference in the \jly between the two studied collision energies increases with \pttrig from about a factor of two at \pttrig~=~2.5~GeV/$c$ to a factor of four at \pttrig~=~5.5~GeV/$c$. Comparisons to \PT simulations are shown as lines in \Fref{Figure1}.  It is surprising how well \PT is able to describe the \pttrig dependence of the \jly in \AplusA collisions.  In general, the agreement is better at larger \pttrig (\pttrig~$>$~4~GeV/$c$), while at lower \pttrig values \PT predicts a larger \jly than observed in the data.  No significant differences between \dAu, \Cu and \Au collisions at \sNNtwohundred are observed.  This finding is consistent with the \jlc arising from fragmentation.

\begin{figure}[t!]
\rotatebox{0}{\resizebox{8.6cm}{!}{
	\includegraphics{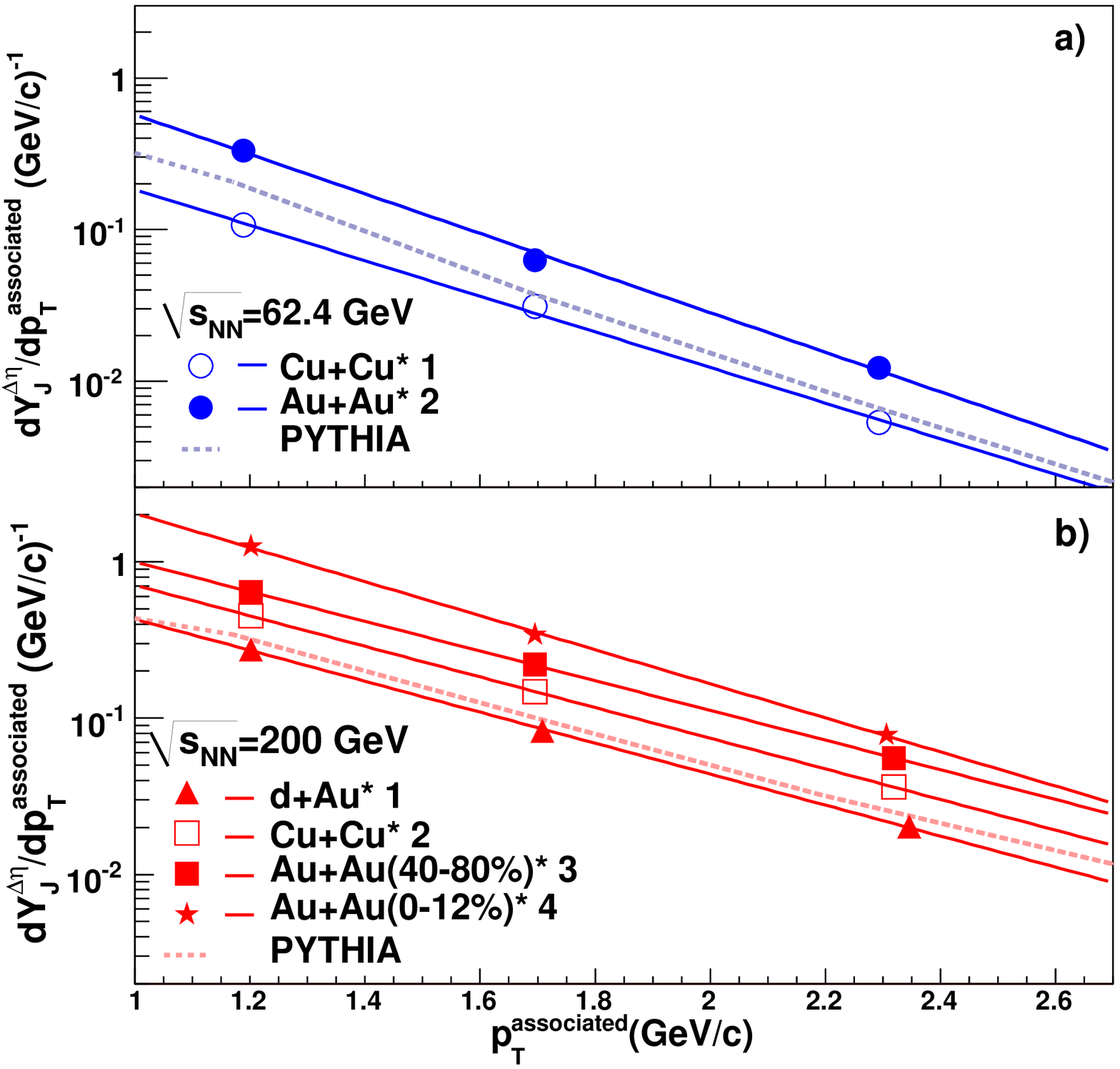}
}}
\caption{(Color online) Dependence of \jly on \ptassoc for \stdtrig for (a) 0-60\% \Cu and 0-80\% \Au collisions at \sNNsixtytwo and (b) 0-95\% \dAu, 0-60\% \Cu, 0-12\% \Au and 40-80\% \Au collisions at \sNNtwohundred.  Solid lines through the data points are fits to the data.  Comparisons to \PT are shown as dashed lines.  The 5\% systematic error due to the uncertainty on the associated particle's efficiency is not shown and systematic errors due to the acceptance correction are given in \Tref{errortable}.}\label{Figure2}
\end{figure}

\begin{figure*}[t!]
\rotatebox{0}{\resizebox{15cm}{!}{
	\includegraphics{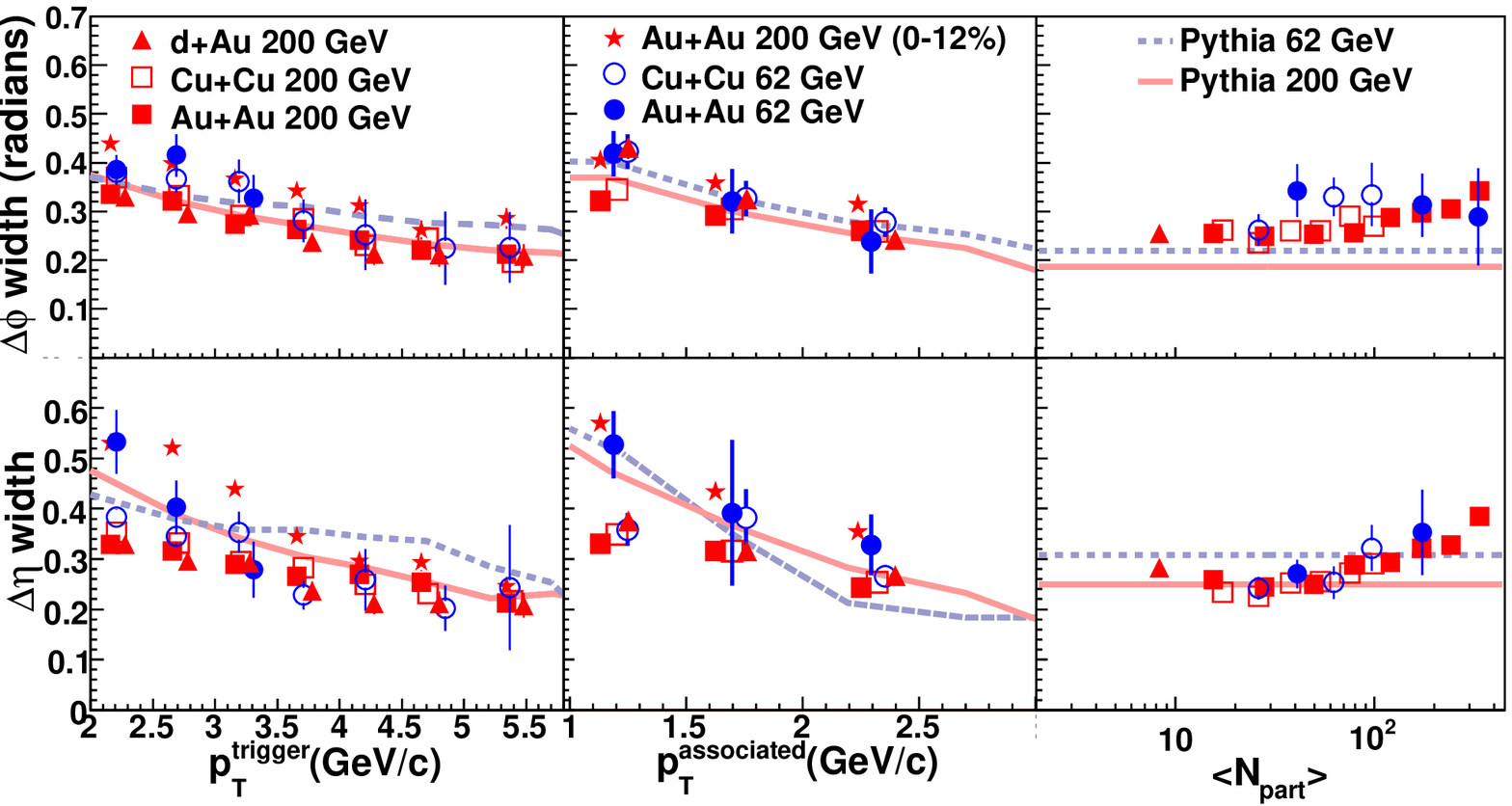}
}}
\caption{(Color online) Dependence of the widths in \dphi and \deta on \pttrig for \stdassoc, \ptassoc for \stdtrig, and \npartav for \stdtrig and \stdassoc for 0-95\% \dAu, 0-60\% \Cu at \sNNsixtytwo and \sNNtwohundred, 0-80\% \Au at \sNNsixtytwo, and 0-12\% and 40-80\% \Au at \sNNtwohundred.  Comparisons to \PT are shown as lines. The 5\% systematic error due to the uncertainty on the acceptance correction is not shown and systematic errors due to the acceptance correction are given in \Tref{errortable}.
  }\label{Width}
\end{figure*}

\begin{table}[b!]
\caption{Inverse slope parameters in \MeV of \ptassoc spectra from fits of data in \Fref{Figure2}.  The inverse slope parameter from a fit to $\pi^-$ inclusive spectra in \Au collisions \cite{Pion,PhysRevLett.97.152301} above 1.0 \GeV is  280.9 $\pm$ 0.4 \MeV for 0-10\% \sNNsixtytwo and 330.9 $\pm$ 0.3 \MeV for 0-12\% \sNNtwohundred.  Statistical errors only.}\label{Table}
\label{Table-slopes}
\begin{tabular}{c|c|c}
\hline
 & \sNNsixtytwo & \sNNtwohundred\\ \hline
\multirow{2}{*}{\Au} & \multirow{2}{*}{332 $\pm$ 13} & 457 $\pm$ 4 (40-80\%)\\
&  & 399 $\pm$  4 (0-12\%)\\
\Cu & 370 $\pm$ 9 & 443 $\pm$ 3 \\ 
\dAu &  & 438 $\pm$ 9 \\ \hline
\PT & 417 $\pm$ 9 & 491 $\pm$ 3 \\ \hline
\end{tabular}
\end{table}
The spectra of particles associated with the \jlc and their comparison to PYTHIA simulations for \stdtrig are shown in \Fref{Figure2}. 
For the same \pttrig selection, the mean transverse momentum fraction $z_T$ carried by the leading hadron is larger at \sNNsixtytwo than at \twohundred due to the steeper jet spectrum. This is reflected in softer \ptassoc spectra at \sNNsixtytwo. The inverse slope parameters from an exponential fit to these data are shown in Table~\ref{Table-slopes}.    There is no difference seen between \Cu and peripheral \Au in either the data points or the extracted inverse slope parameter.  The inverse slope parameter of the central \Au data at \sNNtwohundred is somewhat lower than the other data at \sNNtwohundred, largely because of the larger yield at the lowest \ptassoc.  This also indicates that there is some modification of the \jlc at low \pT.  
While the agreement with \PT is remarkable for a comparison to \AplusA collisions, the discrepancies between \PT and the data are larger at lower momenta and lower energy.
This is expected since \PT is tuned better at higher \pT and higher energy.  

\begin{table*}[t!]
\begin{center}
\begin{tabular}{ccllccccc}
System & $\sigma/\sigma_{\mathrm{geo}}$ & \npartav & \ncollav & $b_{\Delta\phi}$ & $\langle v_2^{\mathrm{trigger}}\rangle$  & $\langle v_2^{\mathrm{assoc}}\rangle$ & $V_{2\Delta}$ & $V_{3\Delta}$ \\[3pt]
 &   [\%] & & & & [\%] &  [\%] & [\%$^2$] & [\%$^2$] \\ \hline\\[-8pt]
\dAu  &  MB &	8.3$\pm$0.4	&	7.5$\pm$0.4     &0.158$\pm$0.006  & 0.0 & 0.0 & 69 $\pm$ 1 & -285 $\pm$ 14 \\[4pt] 
200 GeV   &       &     &                 &  &  & &  &  \\[4pt]  \hline 
\Cu &  0-10 &  96$\pm$3  & 162$\pm$14 &0.885$\pm$      0.010& $8.5_{-2.5}^{+1.4}$& $7.3_{-0.7}^{+1.9}$ & 84 $\pm$ 5 & 54 $\pm$ 5 \\[4pt] 
 62.4 GeV   &	10-30 &  64$\pm$1 & 87.9$\pm$-7.9 &0.515$\pm$      0.006& $11.6_{-3.2}^{+0.5}$& $10.4_{-1.8}^{+0.3}$ & 228 $\pm$ 5 & 5 $\pm$ 5 \\[4pt]        
 &	30-60 &  25.7$\pm$0.6 & 27.6$\pm$1.6 &0.201$\pm$      0.005& $14.5_{-5.3}^{+1.0}$& $11.8_{-2.7}^{+0.3}$  & 471 $\pm$ 10 & -77 $\pm$ 10 \\[2pt]  \hline
\Au &  0-10 & 320$\pm$5  & 800 $\pm$ 74 & 3.582$\pm$0.019& $8.5_{-3.2}^{+0.3}$& $7.6_{-1.9}^{+0.3}$  & 63 $\pm$ 3 &  42 $\pm$ 2 \\[4pt] 
62.4 GeV & 10-40 & 169$\pm$9 & 345 $\pm$ 44 & 1.846$\pm$0.010& $17.5_{-3.2}^{+0.3}$& $15.0_{-1.9}^{+0.3}$  & 258 $\pm$ 2 & 57 $\pm$ 2 \\[4pt] 
  & 40-80 & 42$\pm$8 & 51 $\pm$ 16 & 0.446$\pm$0.009& $21.2_{-5.9}^{+0.3}$& $18.6_{-3.4}^{+0.3}$  & 456 $\pm$ 8 & -33 $\pm$ 8 \\[4pt]  \hline
\Cu &  0-10 &	99$\pm$1   & 189$\pm$15 &1.759$\pm$      0.007& $8.0_{-3.0}^{+0.7}$& $8.8_{-2.2}^{+0.1}$ & 128 $\pm$ 2 & 65 $\pm$ 1 \\[4pt] 
  200 GeV & 10-20 &	75$\pm$1  & 123.6$\pm$9.4  &1.206$\pm$      0.006& $10.5_{-2.5}^{+0.6}$& $11.3_{-1.8}^{+0.0}$ & 212 $\pm$ 2 & 62 $\pm$ 2 \\[4pt] 
      & 20-30 &	54$\pm$1  & 77.6$\pm$5.4 &0.815$\pm$      0.006& $11.2_{-2.9}^{+0.7}$& $12.5_{-1.9}^{+0.1}$ & 296 $\pm$ 3 & 47 $\pm$ 3 \\[4pt] 
      & 30-40 &	38$\pm$1  & 47.7$\pm$2.8 &0.540$\pm$      0.006& $10.7_{-2.6}^{+0.9}$& $13.0_{-2.2}^{+0.1}$ & 380 $\pm$ 4 &  9 $\pm$ 4 \\[4pt] 
      & 40-50 &	26.2$\pm$0.5 & 29.2$\pm$1.6 &0.337$\pm$      0.005& $11.2_{-6.3}^{+2.2}$& $12.6_{-3.7}^{+0.5}$ & 506 $\pm$ 6 &  -37 $\pm$ 6 \\[4pt] 
      & 50-60 &	17.2$\pm$0.4  & 16.8$\pm$0.9 &0.209$\pm$      0.005& $10.3_{-2.7}^{+0.6}$& $12.3_{-2.6}^{+0.0}$ & 657 $\pm$ 10 &  -162 $\pm$ 10 \\[4pt]  \hline
\Au &  0-12 & 	316$\pm$6 &  900$\pm$71 &13.255$\pm$0.003& $8.5_{-2.1}^{+2.1}$& $7.3_{-1.5}^{+1.5}$ & 81 $\pm$ 1 & 49.9 $\pm$ 0.2 \\[4pt]
 200 GeV & 10-20 & 	229$\pm$5 & 511$\pm$34&4.683$\pm$0.007& $14.6_{-2.0}^{+2.0}$& $13.0_{-1.2}^{+1.2}$ & 225 $\pm$ 1 & 67.1 $\pm$ 0.6  \\[4pt]
      & 20-30 & 	164$\pm$5 & 325$\pm$23 &3.222$\pm$0.006& $17.8_{-2.1}^{+2.1}$& $16.5_{-1.2}^{+1.2}$ & 344 $\pm$ 1 & 71.2 $\pm$ 0.8 \\[4pt]
      & 30-40 & 	114$\pm$5 & 199$\pm$16 &2.094$\pm$0.006& $18.9_{-2.3}^{+2.3}$& $18.1_{-1.4}^{+1.4}$ & 430 $\pm$ 1 & 68 $\pm$ 1 \\[4pt]
      & 40-50 & 	76$\pm$5  & 115$\pm$12 &1.267$\pm$0.006& $11.5_{-2.3}^{+2.3}$& $19.0_{-2.5}^{+2.5}$ & 461 $\pm$ 2 & 57 $\pm$ 2 \\[4pt]
      & 50-60 & 	48$\pm$5  & 61$\pm$8 &0.738$\pm$0.006& $10.1_{-2.3}^{+2.3}$& $17.6_{-2.8}^{+2.8}$ & 478 $\pm$ 3 &  18 $\pm$ 3 \\[4pt]
      & 60-70 & 	28$\pm$4  & 30$\pm$5 &0.386$\pm$0.006& $8.5_{-2.2}^{+2.2}$& $15.4_{-2.9}^{+2.9}$ & 549 $\pm$ 6 & -25 $\pm$ 6  \\[4pt]
      & 70-80 & 	15$\pm$2  &  14$\pm$3 &0.183$\pm$0.006& $6.6_{-1.9}^{+1.9}$& $12.8_{-2.8}^{+2.8}$ & 754 $\pm$ 12  & -195 $\pm$ 12 \\[4pt]
\hline
\end{tabular}
\end{center}
\vskip -0.25cm
\caption{
 The background terms $b_{\Delta\phi}$ (see \Eref{Eq:bdef}), elliptic flow values of trigger ($v_2^{\mathrm{trigger}}$) and associated particles ($v_2^{\mathrm{assoc}}$),  and Fourier coefficients $V_{2\Delta}$ and $V_{3\Delta}$ from 2D fits for different collision energies, systems and centrality bins defined by the fraction of geometric cross section ($\sigma/\sigma_{\mathrm{geo}}$), average number of participants (\npartav) and binary collisions (\ncollav) for the data in \Fref{Figure3}.}
\label{tabnpart}
\end{table*}

\Fref{Width} shows the Gaussian widths of the \dphi and \deta projections of the near-side jet-like peak as a function of \pttrig, \ptassoc, and \npartav along with \PT simulations.  In the most central bin in \Au collisions at \sNNsixtytwo it was not possible to extract the width in \deta because of limited statistics combined with a residual track merging effect.  There are no significant differences between the widths as a function of \pttrig and \ptassoc for different collision systems except for central \Au collisions at \sNNtwohundred.  While no dependence on collision system is observed, there is a clear increase in the \deta width with increasing \npartav in \Au collisions at \sNNtwohundred.  This indicates that the shape of the \jlc is modified in central \Au collisions at \sNNtwohundred.  
\PT predicts a greater width in \deta at the lowest \ptassoc than seen in \dAu or \Cu.

Overall it can be concluded that the agreement between the different collision systems and energies shows remarkably little dependence of the jet-like per trigger yield on the system size. In contrast to the peripheral \Au data, the central \Au data show indications that the \jlc is modified.  The model in~\cite{LongFlow}, a hypothesis for the formation of the \ridge through gluon bremsstrahlung,  does not produce a \ridge broad enough to agree with the data, however, it is possible that a similar mechanism could explain the broadening of the \jlc.  Similarly models for \ridge production by turbulent color fields \cite{Romatschke:2006bb,Majumder:2006wi} predict a broadening of the jet-like peak in \deta which is not wide enough to describe the \ridge but may explain the data in \Fref{Width}.

\subsection{The near-side \ridge}

\begin{figure}[b!]
\rotatebox{0}{\resizebox{8.6cm}{!}{
	\includegraphics{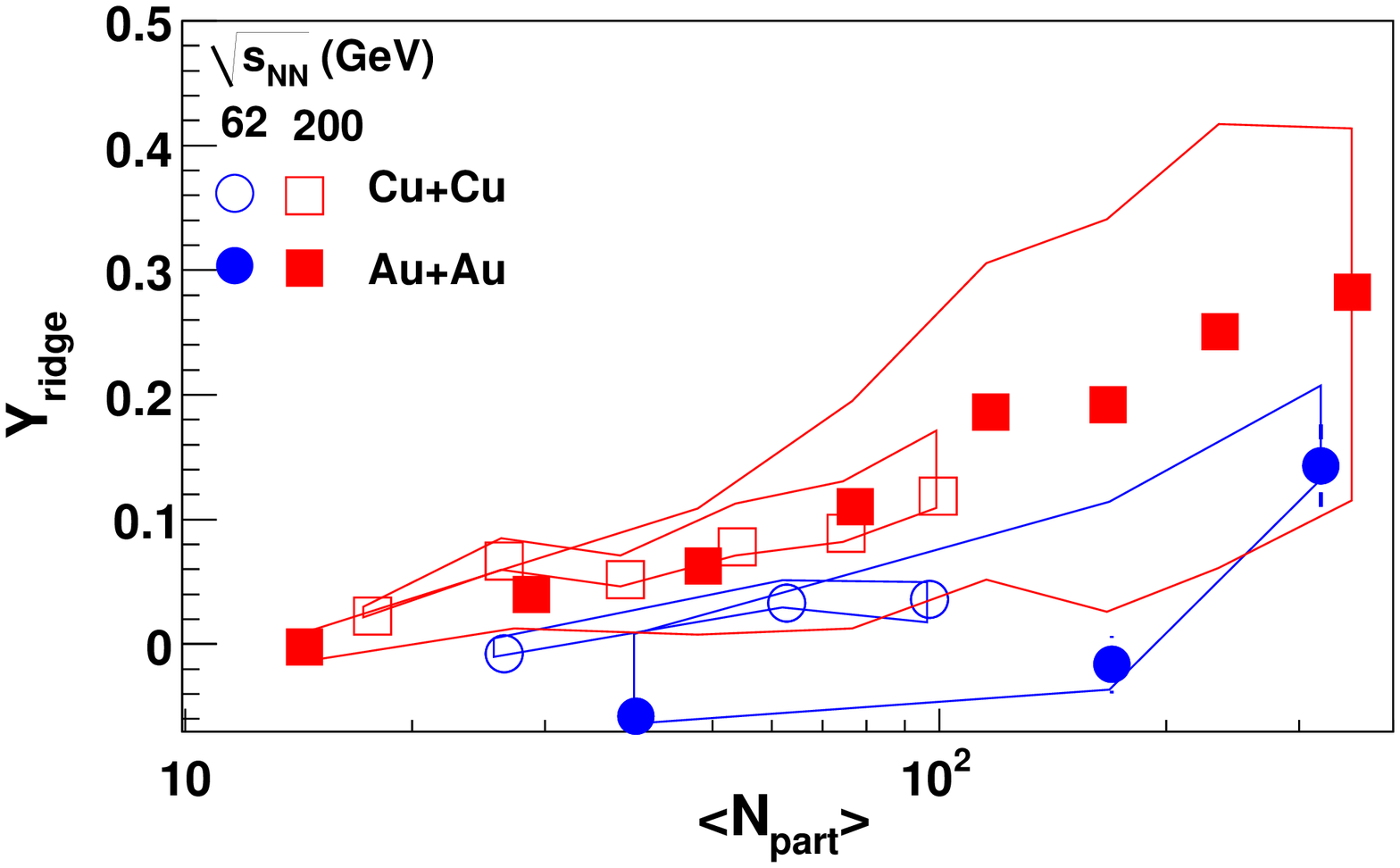}}}
\caption{(Color online) Dependence of the \ridge yield on \npartav for \stdtrig and \stdassoc for \Cu, and \Au at \sNNsixtytwo and \Cu and \Au at \sNNtwohundred.  Systematic errors due to \vtwo are shown as solid lines.  The 5\% systematic error due to the uncertainty on the associated particle's efficiency is not shown and systematic errors due to the acceptance correction are given in \Tref{errortable}.}
\label{Figure5}
\end{figure}

In~\cite{RidgePaper:2009qa}, we reported detailed studies of the \ridge in \Au collisions at \sNNtwohundred as a function of \pttrig and \ptassoc.  Here we investigate the ridge centrality, energy and system size dependence. The dependence of the \ridge yield on \npartav for \Cu and \Au collisions is shown in \Fref{Figure5} for both energies studied.  \Tref{tabnpart} shows $b_{\Delta\phi}$ values and \vtwo of trigger and associated particles for all collision systems and energies studied in \Fref{Figure5}. The centrality bins are characterized by the fraction of geometric cross section $\sigma/\sigma_{\mathrm{geo}}$, average number of participants \npartav and number of binary collisions \ncollav.  Contrary to the \jly, which shows little dependence on centrality, the \ridge yield increases steeply with \npartav.  Within errors, there is no difference in \ridge yield between \Cu and \Au collisions at the same \npartav at a given energy, demonstrating the system independence of the \ridge yield.

\begin{figure}[t!]
\rotatebox{0}{\resizebox{8.6cm}{!}{
	\includegraphics{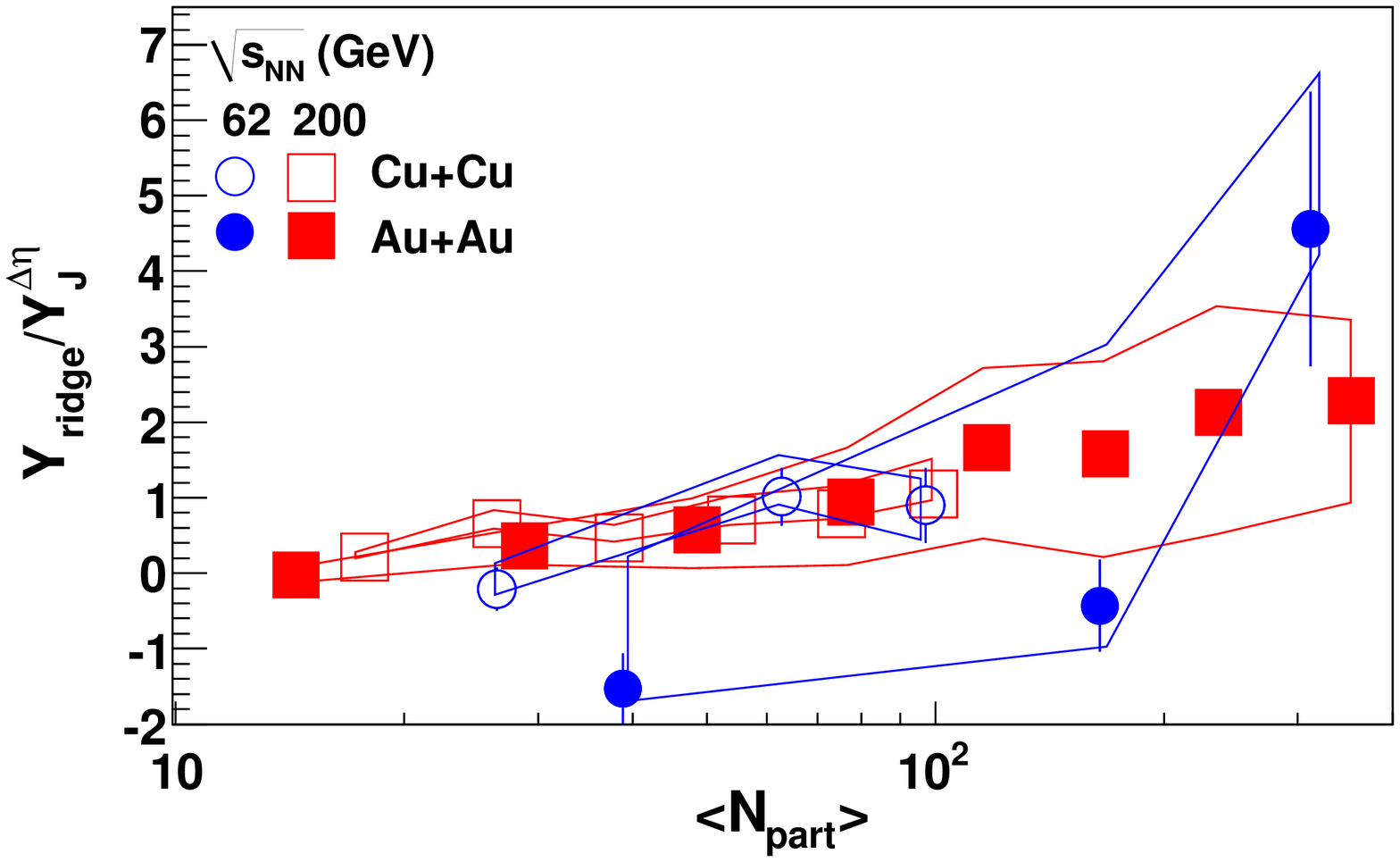}
}}
\caption{(Color online) Ratio of the \ridge and \jlys as a function of \npartav for \stdtrig and \stdassoc.  Systematic errors due to \vtwo are shown as solid lines and systematic errors due to the acceptance correction are given in \Tref{errortable}.}\label{FigureJetRidgeRatio}
\end{figure}

The energy dependence of the ridge yield is potentially a sensitive test of ridge models. Comparing the two collision energies studied, the \ridge yield is observed to be smaller at \sNNsixtytwo than at \sNNtwohundred. Similar behavior was also observed for the \jly. Therefore a closer investigation of the centrality dependence of the ratio $Y_{\mathrm{ridge}}/Y_{\mathrm{jet}}$ is reported in \Fref{FigureJetRidgeRatio}. The ratio of the yields is independent of collision energy within errors. For the same kinematic selections, the data at \sNNsixtytwo correspond to a lower jet energy which may imply the decrease of the ridge yield with the parton energy, as observed for the \jly.

A recent STAR study of the ridge using two-particle azimuthal correlations with respect to the event plane~\cite{Agakishiev:2010ur} in \Au collisions at \sNNtwohundred shows that the ridge yield is dependent on the angle of the trigger particle relative to the event plane.  Another STAR study of three-particle correlations in pseudorapidity~\cite{Abelev:2009jv} in \Au collisions at \sNNtwohundred shows that within current experimental precision particles in the ridge are uncorrelated in $\Delta\eta$ with both the trigger particle and each other.
These observations along with the PHOBOS measurement of the ridge out to \deta=4~\cite{Alver:2009id} provide substantial experimental constraints on theories for the production of the ridge.  In addition, the CMS experiment recently observed the ridge in high multiplicity \pp collisions at \sqrts=7 TeV~\cite{Khachatryan:2010gv}.  We consider these results in addition to the results presented in this paper in order to evaluate our current theoretical understanding of the \ridge.


The momentum kick model~\cite{Wong:2008yh} with the same kinematic selection criteria applied to charged particles describes the increase of the ridge yield with centrality quantitatively in \Au collisions at \sNNtwohundred.  For a given collision energy the ridge yield for \Cu collisions is predicted to approximately follow that for \Au collisions for the same number of participants, however the prediction for \Cu collisions is systematically above that for \Au collisions at both energies.  This difference between the two systems is not corroborated by the data. For the same nucleus-nucleus collisions at different energies, the ridge yield in the momentum kick model is predicted to scale approximately with the number of medium partons produced per participant which increases with increasing collision energy as $(\mathrm{ln}\sqrt{s})^2$~\cite{Busza:2004mc}.  According to this prediction, \nridge should increase by a factor of 1.6 in \sNNtwohundred relative to \sixtytwo.  This is in agreement with our measurement.
We note that the momentum kick model is likely to have difficulty describing the observed dependence of the ridge yield on the event plane, since it would likely predict larger ridge out-of-plane.  It would also likely have difficulty explaining the absence of correlations between particles in the \ridge and the \jlc.

The model where the \ridge arises from the coupling of induced gluon radiation to the longitudinal flow~\cite{Armesto:2004pt} is in qualitative agreement with the observed increase of \nridge as a function of \npartav since the size of the \ridge should depend roughly on the average path length traveled by a hard parton.  However, it is not obvious that this model can describe the collision system and energy dependence of ridge yield reported in this paper. Moreover, the large extent of the ridge in \deta and absence of correlation among ridge particles clearly disfavors this physics mechanism for ridge formation.

The radial flow plus trigger bias model ~\cite{Pruneau:2007ua,Shuryak:2007fu,Voloshin:2003ud} would predict an increasing \ridge yield with increasing \npart.  This model would predict a larger \ridge in-plane because of the larger surface area in-plane, in agreement with the data. Since the ridge in this model arises from medium partons, particles in the \ridge are not expected to be correlated with each other, again in agreement with the 3-particle \deta correlation data. 

\begin{figure*}
\rotatebox{0}{\resizebox{18cm}{!}{
	\includegraphics{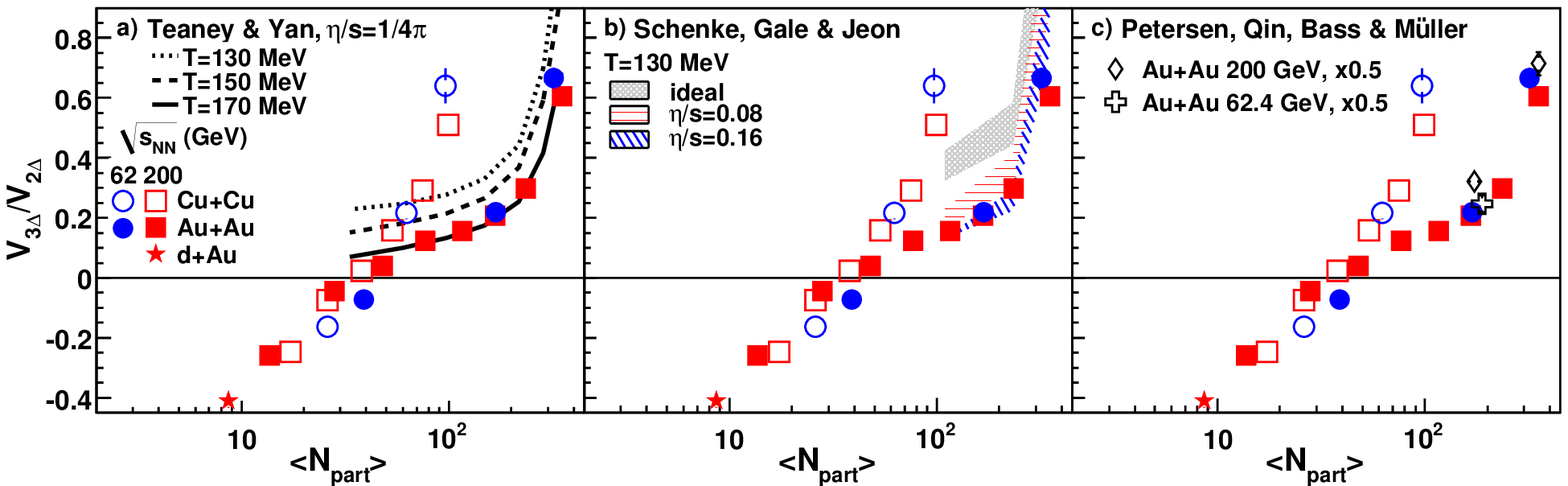}
}}
\caption{(Color online) \vthreeratio ratio as a function of \npartav for \stdtrig and \stdassoc. Statistical errors only.  The systematic error of $<$4\% not shown.  The data are compared to hydrodynamical calculations using (a) lattice equation of state with three different freeze-out temperatures in \Au collisions at \sNNtwohundred from~\cite{Teaney:2010vd,TeaneyPrivateComm} (b) using a 3+1D viscous hydrodynamical model with three different viscosity to entropy ratios, $\eta/s$~\cite{Schenke:2010rr,SchenkePrivateComm}, and (c) a 3+1D model incorporating both hydrodynamics and transport model~\cite{Petersen:2010cw,PetersenPrivateComm}.  (a) and (c) use the same kinematic cuts as the data while (b) uses 1.5~$<$ \pT $<$~4.0 GeV/$c$ for both \pttrig and \ptassoc.}\label{Figurev3Npart}
\end{figure*}
The mechanisms for the production of the \ridge through glasma initial state effects~\cite{Dumitru:2008wn,Glasma2,Gavin:2008ev} are able to explain the observed large $\Delta\eta$ extent of the \ridge. However, it is not obvious what this class of models would predict for the collision system and energy dependence discussed in this paper as well as for other ridge properties including its \pttrig and \ptassoc dependence, the event plane dependence, and the absence of correlated structures in three-particle \deta correlations.    While there are calculations of untriggered di-hadron correlations for the glasma model, there are no calculations to compare to \highpT triggered correlations.  It should be noted that these calculations not only include the glasma initial state but also hydrodynamical flow so some of the \ridge in these models is created by flow.  
If the ridge is produced by the same mechanism in \pp and \AplusA collisions, models where the ridge is produced by initial state effects such as the glasma model may be the only models able to explain both the \pp and \AplusA data simultaneously since hydrodynamical flow in \pp collisions is expected to be small if not negligible. Therefore quantitative calculations of the \ridge in this class of models and the identified particle spectra measurements in  high multiplicity \pp collisions are essential for understanding the production mechanism of the \ridge.

Models describing the ridge in terms of quadrupole, triangular and higher order components, $v_{\mathrm{n}}$, from initial eccentricity fluctuations predict a ridge yield that increases with \npart~\cite{Sorensen:2010zq,Alver:2010gr,Alver:2010grErratum,Sorensen:2011xw} in qualitative agreement with data. 
Motivated by these models, we applied two dimensional fits to our data with a two dimensional Gaussian to describe the jet-like component and \deta independent $V_{\mathrm{n}\Delta}$ terms given by Eq.~(\ref{eq:2dfit}). \Fref{Figurev3Npart} shows \vthreeratio as a function of \npartav from these fits.  The values of the Fourier coefficients \vthreesq and \vtwotildesq are given in \Tref{tabnpart}.  We allow \vthreesq to be negative and \vthreesq is negative for \dAu collisions and peripheral \AplusA collisions.  An approximately Gaussian peak from an \as jet-like correlation would give a negative \vthreesq and indicate that \vthreesq is dominantly from non-flow.  The ratio \vthreeratio evolves from negative values in \dAu and peripheral \AplusA collisions to positive values at larger \npartav.  

\noindent Positive values for \vthreesq are consistent with expectations from triangular initial conditions.  Contributions to a jet-like peak on the \as, such as that observed in \dAu, would lead to \vthreeratio $<$ \vthreeflowratio because \vthreesq would be an underestimate of $v_3^2$ and \vtwotildesq would be an overestimate of $v_2^2$. For both \Au and \Cu collisions, the ratio is independent of collision energy within errors and is largest in the most central collisions with significant deviation between the two colliding systems at the same \npartav.
It is not clear at this point whether the representation of the data in \Fref{Figure5}, where the \vtwo has been subtracted, or in \Fref{Figurev3Npart}, where the \vtwo is explicitly included, gives the most insight into the production mechanism of the \ridge.

We compare the data to three hydrodynamical models in \Fref{Figurev3Npart}, noting that these data are in a momentum range approaching the limit where hydrodynamical models are expected to be valid.  By comparing different models to the data we are able to see whether the data can constrain the initial state.
In \Fref{Figurev3Npart}a we compare the data to 2+1D hydrodynamical model calculations for \Au collisions at \sNNtwohundred for various hadron resonance gas freezeout temperatures and using a modified Glauber initial state~\cite{Teaney:2010vd}.  These calculations use a lattice equation of state and a viscosity to entropy ratio $\eta$/s = 1/4$\pi$~\cite{TeaneyPrivateComm}.  We note that this model does not include resonance decays.  These predictions agree with the data for all but the three most peripheral centrality bins for a hadron gas freezeout temperature of 170~MeV.

In \Fref{Figurev3Npart}b we compare the data to 3+1D hydrodynamical model calculations for several values of $\eta$/s and a kinetic freezeout temperature of 130 MeV~\cite{Schenke:2010rr,SchenkePrivateComm}.  This model uses a Glauber initial state modified to generate structures like those expected from flux tubes and differs from ~\cite{Teaney:2010vd} because it includes nontrivial longitudinal dynamics and the effect of varying the $\eta$/s at a given freeze-out temperature. The calculations in~\cite{Schenke:2010rr,SchenkePrivateComm} were limited in momentum so it was only possible to compare to calculations with momenta of 1.5~$<$ \pT $<$~4.0 GeV/$c$ for both \pttrig and \ptassoc. Qualitatively this model reproduces the trends observed in the data and it tends to support a higher value of $\eta$/s.%

\Fref{Figurev3Npart}c shows a comparison of the data to a 3+1D event-by-event transport + hydrodynamical model calculation~\cite{Petersen:2010cw,PetersenPrivateComm}. In this model, the initial conditions such as long range rapidity correlations and fluctuations in the transverse energy density profile are provided by the UrQMD model~\cite{Bass:1998ca,Bleicher:1999xi}. The hydrodynamic evolution starts at 0.5 fm. A transition from hydrodynamic evolution to the transport approach is followed by final state rescatterings and resonance decays. The predictions shown in the figure are scaled by a factor 0.5.  Requiring a high $p_T$ trigger particle skews the calculation towards events with a hot spot in the initial density profile, which may lead to a preference for events with a high $v_3$ value.  This model qualitatively describes the \npartav dependence of \vthreeratio for \Au collisions at both \sNNsixtytwo and \twohundred. %

Despite differences in initial conditions, transport parameters and freeze-out requirements, all three models shown are able to reproduce the qualitative trends of \vthreeratio versus \npartav for large \npartav. Since $V_{3\Delta}$ is expected to be largely independent of centrality~\cite{Qiu:2011iv,Alver:2010gr} this likely reflects the model's accuracy in predicting centrality dependence of $V_{2\Delta}$. The models in~\cite{TeaneyPrivateComm,Schenke:2010rr} reproduce the data fairly well quantitatively, but ~\cite{TeaneyPrivateComm} achieves the best fit by varying the temperature while~\cite{Schenke:2010rr} varies the viscosity. Since they also have different initial conditions, we infer that agreement with the data is possible even with different assumptions and parameters during the hydrodynamical evolution.  It will be interesting to see if such similarity persists at lower transverse momenta where the hydrodynamic calculations are more reliable.  These studies imply that while fluctuations of the initial state are needed to induce odd higher order $v_n$ terms, the observable \vthreeratio is rather insensitive to the exact details of the model. This measurement does however enforce added restrictions to model implementations and should therefore be added to the suite of results, such as identified particle spectra, yields and \pT dependence of \vtwo, currently used to validate theories. Future theoretical and experimental studies will be needed to determine whether these models are sufficient to describe the complete \ridge, and/or whether there are substantial non-flow contributions to the Fourier coefficient $V_{3\Delta}$ in central \AplusA collisions.

\section{Conclusions}
The energy and system size dependence of \ns di-hadron correlations enables studies of the \jlc and the \ridge at fixed densities with different geometry.
The reasonable agreement of the \jlc with \PT is surprising, especially considering that \PT is a \pp event generator.  There is remarkably little dependence on the collision system at both \sNNsixtytwo and \twohundred except for central \Au collisions at \sNNtwohundred.  In central \Au collisions at \sNNtwohundred the \jlc is substantially broader and the spectrum softer than in peripheral collisions and than those in collisions of other systems in this kinematic regime.  This may indicate that fragmentation is modified in these collisions so that the parton fragments softer, perhaps due to a mechanism such as gluon bremsstrahlung.  This indicates that the near-side \jlc is dominantly produced by vacuum fragmentation.

The \ridge is observed not only in \Au collisions at \sNNtwohundred as we reported earlier, but also in \Cu collisions and in both studied collision systems at lower energy of \sNNsixtytwo.  This demonstrates that the \ridge is not a feature unique to \Au collisions at the top RHIC energy.  
We observe two trends which set significant limits to models.  
First, when the \ridge is measured using the standard ZYAM model, the \ridge is comparable in \Cu and \Au collisions and the energy dependence of the \ridge is the same as the energy dependence of the \jlc.
Second, when we subtract the \jlc and calculate the third component of the Fourier decomposition \vthree, which is $v_3^2$ in the absence of non-flow, we see different trends for \Cu and \Au but no difference between the two energies.
The combination of these data with future measurements at lower RHIC energies ($\sqrt{s_{{NN}}}$~=~7-39~GeV) as well as studies at the LHC will therefore be a powerful tool for the distinction between various theoretical models for the production of the \ridge.


\vspace{1.0cm}

\begin{acknowledgments}

 We thank Hannah Petersen, Bjoern Schenke, Derek Teaney and Li Yan for useful discussions, the RHIC Operations Group and RCF at BNL, the NERSC Center at 
LBNL and the 
Open Science Grid consortium for providing resources and support. This work was supported in part by the Offices of NP and HEP within the U.S. DOE Office of Science, the U.S. NSF, the Sloan Foundation, the DFG cluster of excellence `Origin and Structure of the Universe'of Germany, CNRS/IN2P3, FAPESP CNPq of Brazil, Ministry of Ed. and Sci. of the Russian Federation, NNSFC, CAS, MoST, and MoE of China, GA and MSMT of the Czech Republic, FOM and NWO of the Netherlands, DAE, DST, and CSIR of India, Polish Ministry of Sci. and Higher Ed., Korea Research Foundation, Ministry of Sci., Ed. and Sports of the Rep. Of Croatia, and RosAtom of Russia.

\end{acknowledgments}
\bibliography{Bibliography}

\end{document}

%% file: authorList.tex
\affiliation{Argonne National Laboratory, Argonne, Illinois 60439, USA}
\affiliation{Brookhaven National Laboratory, Upton, New York 11973, USA}
\affiliation{University of California, Berkeley, California 94720, USA}
\affiliation{University of California, Davis, California 95616, USA}
\affiliation{University of California, Los Angeles, California 90095, USA}
\affiliation{Universidade Estadual de Campinas, Sao Paulo, Brazil}
\affiliation{University of Illinois at Chicago, Chicago, Illinois 60607, USA}
\affiliation{Creighton University, Omaha, Nebraska 68178, USA}
\affiliation{Czech Technical University in Prague, FNSPE, Prague, 115 19, Czech Republic}
\affiliation{Nuclear Physics Institute AS CR, 250 68 \v{R}e\v{z}/Prague, Czech Republic}
\affiliation{University of Frankfurt, Frankfurt, Germany}
\affiliation{Institute of Physics, Bhubaneswar 751005, India}
\affiliation{Indian Institute of Technology, Mumbai, India}
\affiliation{Indiana University, Bloomington, Indiana 47408, USA}
\affiliation{Alikhanov Institute for Theoretical and Experimental Physics, Moscow, Russia}
\affiliation{University of Jammu, Jammu 180001, India}
\affiliation{Joint Institute for Nuclear Research, Dubna, 141 980, Russia}
\affiliation{Kent State University, Kent, Ohio 44242, USA}
\affiliation{University of Kentucky, Lexington, Kentucky, 40506-0055, USA}
\affiliation{Institute of Modern Physics, Lanzhou, China}
\affiliation{Lawrence Berkeley National Laboratory, Berkeley, California 94720, USA}
\affiliation{Massachusetts Institute of Technology, Cambridge, MA 02139-4307, USA}
\affiliation{Max-Planck-Institut f\"ur Physik, Munich, Germany}
\affiliation{Michigan State University, East Lansing, Michigan 48824, USA}
\affiliation{Moscow Engineering Physics Institute, Moscow Russia}
\affiliation{NIKHEF and Utrecht University, Amsterdam, The Netherlands}
\affiliation{Ohio State University, Columbus, Ohio 43210, USA}
\affiliation{Old Dominion University, Norfolk, VA, 23529, USA}
\affiliation{Panjab University, Chandigarh 160014, India}
\affiliation{Pennsylvania State University, University Park, Pennsylvania 16802, USA}
\affiliation{Institute of High Energy Physics, Protvino, Russia}
\affiliation{Purdue University, West Lafayette, Indiana 47907, USA}
\affiliation{Pusan National University, Pusan, Republic of Korea}
\affiliation{University of Rajasthan, Jaipur 302004, India}
\affiliation{Rice University, Houston, Texas 77251, USA}
\affiliation{Universidade de Sao Paulo, Sao Paulo, Brazil}
\affiliation{University of Science \& Technology of China, Hefei 230026, China}
\affiliation{Shandong University, Jinan, Shandong 250100, China}
\affiliation{Shanghai Institute of Applied Physics, Shanghai 201800, China}
\affiliation{SUBATECH, Nantes, France}
\affiliation{Texas A\&M University, College Station, Texas 77843, USA}
\affiliation{University of Texas, Austin, Texas 78712, USA}
\affiliation{University of Houston, Houston, TX, 77204, USA}
\affiliation{Tsinghua University, Beijing 100084, China}
\affiliation{United States Naval Academy, Annapolis, MD 21402, USA}
\affiliation{Valparaiso University, Valparaiso, Indiana 46383, USA}
\affiliation{Variable Energy Cyclotron Centre, Kolkata 700064, India}
\affiliation{Warsaw University of Technology, Warsaw, Poland}
\affiliation{University of Washington, Seattle, Washington 98195, USA}
\affiliation{Wayne State University, Detroit, Michigan 48201, USA}
\affiliation{Institute of Particle Physics, CCNU (HZNU), Wuhan 430079, China}
\affiliation{Yale University, New Haven, Connecticut 06520, USA}
\affiliation{University of Zagreb, Zagreb, HR-10002, Croatia}

\author{G.~Agakishiev}\affiliation{Joint Institute for Nuclear Research, Dubna, 141 980, Russia}
\author{M.~M.~Aggarwal}\affiliation{Panjab University, Chandigarh 160014, India}
\author{Z.~Ahammed}\affiliation{Variable Energy Cyclotron Centre, Kolkata 700064, India}
\author{A.~V.~Alakhverdyants}\affiliation{Joint Institute for Nuclear Research, Dubna, 141 980, Russia}
\author{I.~Alekseev}\affiliation{Alikhanov Institute for Theoretical and Experimental Physics, Moscow, Russia}
\author{J.~Alford}\affiliation{Kent State University, Kent, Ohio 44242, USA}
\author{B.~D.~Anderson}\affiliation{Kent State University, Kent, Ohio 44242, USA}
\author{C.~D.~Anson}\affiliation{Ohio State University, Columbus, Ohio 43210, USA}
\author{D.~Arkhipkin}\affiliation{Brookhaven National Laboratory, Upton, New York 11973, USA}
\author{G.~S.~Averichev}\affiliation{Joint Institute for Nuclear Research, Dubna, 141 980, Russia}
% ...

\collaboration{STAR Collaboration}\noaffiliation